\documentclass[sigconf, nonacm]{acmart}

\newcommand\vldbdoi{XX.XX/XXX.XX}

\newcommand\vldbavailabilityurl{https://github.com/iDC-NEU/DGAI}
\newcommand\vldbpagestyle{plain} 

\usepackage{subcaption}
\usepackage{endnotes,microtype,xspace,graphicx,fancyvrb,multirow} 
\usepackage[linesnumbered, ruled,resetcount, noend]{algorithm2e} 
\usepackage{algpseudocode}
\usepackage{booktabs} 
\usepackage[normalem]{ulem}

\newcommand{\ljh}[1]{\textcolor{orange}{#1}}

\newcommand{\red}[1]{{#1}}

\newcommand{\eat}[1]{}

\usepackage{balance}
\usepackage{latexsym}
\usepackage{amsfonts}
\usepackage{amsmath}
\usepackage{amssymb}
\usepackage{color}
\usepackage{epsfig}
\usepackage{xspace}
\usepackage{graphicx}
\usepackage{cleveref}
\usepackage{balance}
\usepackage{hhline}
\usepackage{float}
\usepackage{xcolor}

\usepackage{epsfig}
\usepackage{multirow}
\usepackage{url}

\usepackage{enumitem}
\setlist{topsep=0pt,noitemsep} \setitemize[1]{label=$\circ$}

\newcommand{\ei}{\end{itemize}\vspace{1ex}}

\newcommand{\ee}{\end{enumerate}\vspace{1ex}}
\newcommand{\beqn}{\begin{eqnarray}}
\newcommand{\eeqn}{\end{eqnarray}}

\newcommand{\stitle}[1]{\vspace{1.2ex}\noindent{\bf #1}}
\newcommand{\etitle}[1]{\vspace{1.2ex}\noindent{\em\uline{#1}}}

\renewcommand{\t}{\tau}

\newcommand{\ie}{i.e.,\xspace}
\newcommand{\eg}{e.g.,\xspace}

\renewcommand{\smallskip}{\vspace{0.6ex}}

\newcounter{ccc}

\newcommand{\eop}{\hspace*{\fill}\mbox{$\Box$}}     
\newcounter{example}
\renewcommand{\theexample}{\arabic{example}}

\newlist{myitemize}{itemize}{3}
\setlist[myitemize,1]{label=$\circ$,leftmargin=3.8ex}
\setlist[myitemize,2]{label=$\bullet$,leftmargin=3.8ex}
\setlist[myitemize,3]{label=$\diamond$, leftmargin=3.2ex}

\newcommand{\mei}{\end{myitemize}\vspace{0.6ex}}

\usepackage{tikz}

\usepackage{comment}
\newcommand{\gsf}[1]{\textcolor{blue}{#1}}

\newcommand{\oursys}{\texttt{DGAI}\xspace}

\begin{document}
\title{DGAI: Decoupled On-Disk Graph‑Based ANN Index for Efficient Updates and Queries}

\settopmatter{authorsperrow=3}

\author{Jiahao Lou}
\affiliation{
  \institution{Northeastern Univ., China}
}
\email{loujh@mails.neu.edu.cn}

\author{Shufeng Gong}
\authornote{Corresponding author: Shufeng Gong (gongsf@mail.neu.edu.cn)}
\affiliation{
  \institution{Northeastern Univ., China}
}
\email{gongsf@mail.neu.edu.cn}

\author{Quan Yu}
\affiliation{
  \institution{Northeastern Univ., China}
}
\email{yuquan@mails.neu.edu.cn}

\author{Hao Guo}
\affiliation{
  \institution{Tsinghua Univ., China}
}
\email{gh23@mails.tsinghua.edu.cn}

\author{Youyou Lu}
\affiliation{
  \institution{Tsinghua Univ., China}
}
\email{luyouyou@tsinghua.edu.cn}

\author{Song Yu}
\affiliation{
  \institution{Northeastern Univ., China}
}
\email{yusong@stumail.neu.edu.cn}

\author{Yanfeng Zhang}
\affiliation{
  \institution{Northeastern Univ., China}
}
\email{zhangyf@mail.neu.edu.cn}

\author{Tiezheng Nie}
\affiliation{
  \institution{Northeastern Univ., China}
}
\email{nietiezheng@mail.neu.edu.cn}

\author{Ge Yu}
\affiliation{
  \institution{Northeastern Univ., China}
}
\email{yuge@mail.neu.edu.cn}

\begin{abstract}
On-disk graph-based indexes are favored for billion-scale Approximate Nearest Neighbor Search (ANNS) due to their high performance and cost-efficiency. However, existing systems typically rely on a coupled storage architecture that co-locates vectors and graph topology, which introduces substantial redundant I/O during index updates, thereby degrading usability in dynamic workloads.

In this paper, we propose a \textit{decoupled storage architecture} that physically separates heavy vectors from the lightweight graph topology.
This design substantially improves update performance by reducing redundant I/O during updates.
However, it introduces I/O amplification during ANNS, leading to degraded query efficiency.
To improve query performance within the update-friendly architecture, we propose two techniques co-designed with the decoupled storage.
We develop a similarity-aware dynamic layout that optimizes data placement online so that redundantly fetched data can be reused in subsequent search steps, effectively turning read amplification into useful prefetching.
In addition, we propose a two-stage query mechanism enhanced by hierarchical PQ, which uses hierarchical PQ to rapidly and accurately identify promising candidates and performs exact refinement on raw vectors for only a small number of candidates. This design significantly reduces both the I/O and computational cost of the refinement stage.
Overall, \oursys achieves resource-efficient updates and low-latency queries simultaneously.
Experimental results demonstrate that \oursys improves update speed by 8.17$\times$ for insertions and 8.16$\times$ for deletions, while reducing peak query latency under mixed workloads by 67\% compared to state-of-the-art baselines.
\end{abstract} 

\maketitle

\pagestyle{\vldbpagestyle}

\ifdefempty{\vldbavailabilityurl}{}{
\vspace{.3cm}
\begingroup\small\noindent\raggedright\textbf{PVLDB Artifact Availability:}\\
The source code, data, and/or other artifacts have been made available at \url{\vldbavailabilityurl}.
\endgroup
}

\section{Introduction}
\label{sec:intro}


Approximate nearest neighbor search (ANNS) plays a critical role in various real-world applications, such as information retrieval~\cite{AsaiMZC23,WilliamsLKWSG14,MohoneyPCMIMPR23,SanthanamKSPZ22}, recommendation systems~\cite{MengDYCLGLC20,SarwarKKR01,CovingtonAS16,OkuraTOT17}, and large language models (LLMs)~\cite{abs-2304-11060,abs-2312-10997,LewisPPPKGKLYR020,abs-2202-01110,jing2024large,bang2023gptcache,abs-2411-05276,VELO,abs-2502-03771}.
Graph-based indices, such as HNSW~\cite{HNSW} and Vamana~\cite{DiskANN}, are widely used to enable efficient ANNS on high-dimensional vectors.
In these indices, vectors are organized as a directed graph, \eat{where vectors are treated as nodes and connected based on their distance relationships.}{where each node corresponds to a vector, and edges are established according to distance relationships.}
To support larger high-dimensional vector datasets, indices are typically stored on disk \cite{FreshDiskANN}, as they are difficult to fit entirely into the memory of a commodity PC~\cite{Starling}.

\eat{
Most existing graph-based indices~\cite{DiskANN,HNSW,NSG,NSW}, along with their algorithm-level~\cite{VBASE,Starling} and system-level optimizations~\cite{osdi25pipeann, yu2022gpu,zhao2020song}, primarily focus on query efficiency, while overlooking the dynamic nature of vector datasets in real scenarios, which frequently evolve through the addition of new vectors and the deletion of outdated ones over time.
For instance, in LLM-based autonomous agents equipped with long-term memory~\cite{xu2025mem,fang2025lightmem,li2025memos,yan2025memory}, the system relies on retrieving relevant past experiences or tools to guide current decision-making. As the agent continuously interacts with the environment, its memory stream evolves rapidly: new observations are embedded and appended, while outdated context is pruned. Consequently, it is critical to support rapid and lightweight updates to maintain index freshness, while preserving the performance of retrieval queries.
}
\eat{
For instance, in e-commerce platforms like Amazon~\cite{Amazon} that support image-based product search, users can find similar products by uploading an image. The system then encodes this image into a vector and performs ANNS to retrieve similar products from the database.
As products change frequently, with new items being added and sold-out items being removed, the vector set evolves over time. Subsequently, the index should be 
updated to maintain query efficiency and accuracy.
}

\begin{figure}[tbp]
        
		\centering
            \includegraphics[width=1\linewidth]{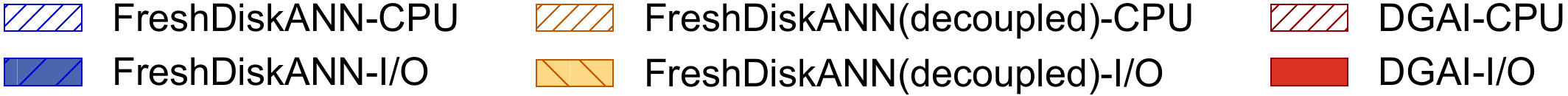}
		\subfloat[Update time breakdown]{
            \label{fig:update_composition}
		  \includegraphics[width=0.47\linewidth]{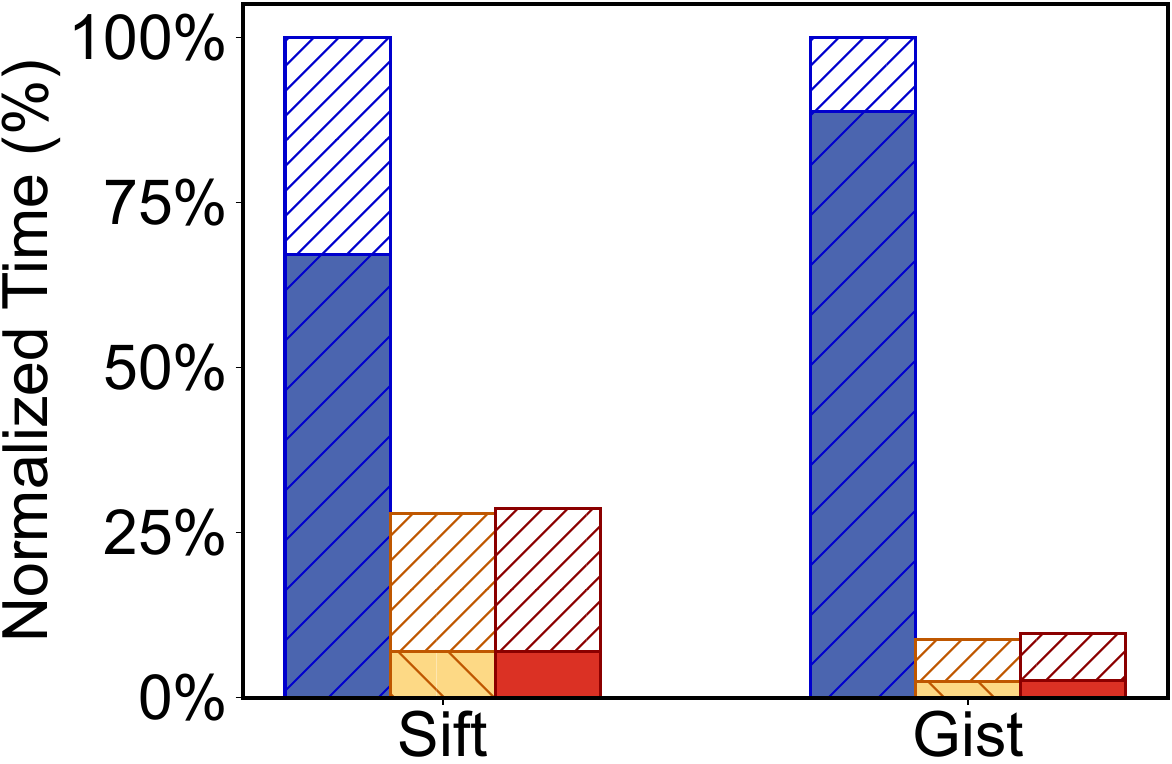}
        }
		\subfloat[Query time breakdown]{
            \label{fig:query_composition}
            \includegraphics[width=0.47\linewidth]{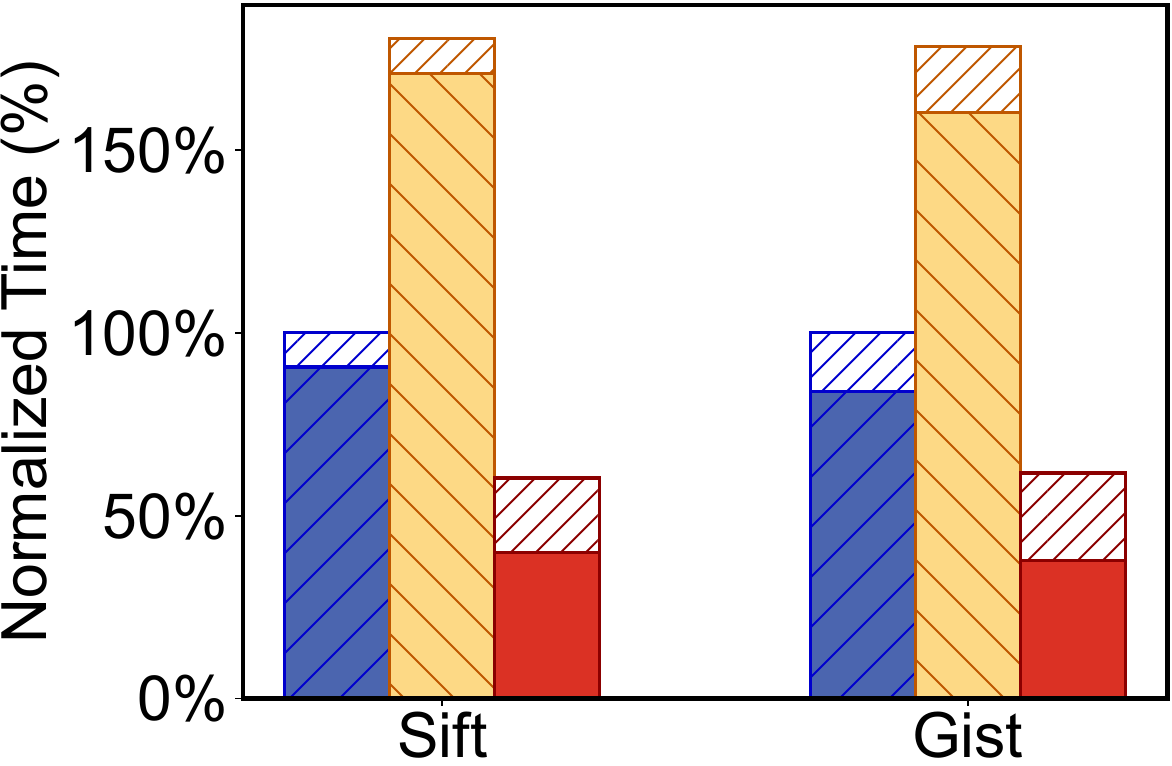}
        }
		\vspace{-0.05in}
        \caption{The update and \eat{ANNS} query time (breakdown) comparison, where the update only contains deletions. The -CPU and -I/O mean the time spent on CPU and disk I/O.
        \eat{(a) Update time breakdown for FreshDiskANN, a naive decoupled design, and our design (\oursys).
        (b) Time breakdown during queries for the three systems.}
        }
        \label{fig:couple_decouple}
         \vspace{-0.2in}
\end{figure}

Most existing graph-based indices~\cite{DiskANN,HNSW,NSG,NSW}, along with various subsequent optimization techniques~\cite{osdi25pipeann,pageann,Starling, zhou2025govector}, primarily focus on query efficiency, while overlooking the \emph{dynamic} scenarios in which new vectors are added to the vector dataset and outdated vectors are deleted from it.
Driven by the surging demand for real-time data freshness, several dynamic ANNS systems have been proposed to support these dynamic evolving workloads~\cite{FreshDiskANN, pgvector,IP-DiskANN,fast26odinann,liuwolverine}.

However, \textit{our experimental analysis reveals that existing graph-based dynamic ANNS systems usually exhibit poor update performance due to high I/O overhead}.
As shown in Figure~\ref{fig:update_composition}, I/O operations dominate the update latency, accounting for \red{57.9–80.5\%} of the total update time.
The issue stems from the \textit{coupled storage architecture}, in which each vector is stored together with its neighbor IDs (\ie graph topology).
This design is well suited for search: graph traversal repeatedly selects the most promising node for expansion, \ie examines its neighbors to identify better candidates, and computes the exact distance to the query for final result ranking. This process requires both neighbor IDs and raw vector, which can be fetched together in a single page under the coupled storage.
While this coupled layout benefits search, it becomes problematic for updates.
An update to a vector involves the insertion or deletion of a corresponding node in the graph topology, which in turn propagates updates to tens or hundreds of neighbors,
requiring the insertion or removal of their incident edges. Under a coupled storage architecture, such topology updates are accompanied by a large number of high-dimensional vector read and write operations, thereby incurring substantial I/O overhead. Profiling reveals that such redundant writes induce over 79\% I/O overhead.
\eat{
Since processing a vector update requires modifying the edges between the updated node and tens of its neighbors, the coupled storage design inevitably incurs redundant rewrites of their large vector data, even though the update only affects the graph topology.
Profiling reveals that such redundant writes induce over 79\% I/O overhead, shifting the bottleneck from compute to I/O.}


\eat{However, we find that more than \red{80\%} of the data involved in I/O operations is unnecessary for updating the index (Section~\ref{sec:motivation:limit}).
Specifically, inserting or deleting a vector requires only a single I/O operation to update the vector data on disk. While updating the graph topology involves modifying the nodes and their edges affected by the deleted or inserted nodes, incuring more I/O operations and overhead than the vector dataset.

\stitle{Motivation}.
Existing graph-based indexes adopt a tightly coupled storage architecture of graph topology and vector for query efficiency, where each node stores its corresponding high-dimensional vector, as illustrated in Figure~\ref{fig:Coupled_index–vector}. 
As a result, index updates involve concurrent read and write operations on the associated vectors, introducing substantial unnecessary I/O overhead. Such inefficiency motivates our design of the decoupled storage architecture.}

\stitle{Decoupled Architecture.} 
In this paper, we adopt a decoupled architecture, \ie \textit{separating the vector from the graph topology}.
Specifically, vectors and topology are maintained as two independent files, enabling independent updates. Compared to the coupled design, this architecture eliminates redundant vector I/O during topology updates, thereby reducing update overhead.
As shown in Figure~\ref{fig:update_composition}, FreshDiskANN \eat{under} with the decoupled design reduces the total update time by 56\% to 84\% compared to the coupled baseline.

\stitle{Dilemma: Update-Efficiency vs. Query-Latency.}
However, although a decoupled architecture improves update throughput, it simultaneously degrades query performance. As shown in Figure~\ref{fig:query_composition}, the decoupled architecture increases query latency by more than 78\%. 
This is mainly caused by the following two reasons:

\etitle{(1) Read amplification.} Decoupled storage separates raw vectors and graph topology, leaving each fetched page with less useful data. As node degree is typically capped at a small constant (e.g., $< 100$), a node’s neighbor IDs occupy at most 400 bytes, far smaller than a 4 KB SSD page, leading to roughly 10$\times$ read amplification.


\etitle{(2) Double read operations.} After decoupling, vectors and graph topology are stored in two separate files, turning what was previously a single read in coupled storage into two separate reads, thereby doubling the I/O cost.

\stitle{Challenges.}
Eliminating these I/O inefficiencies is challenging in a dynamic decoupled index.

\etitle{(1) Maintaining locality under updates.} Read amplification can be reduced by adjusting the layout so that nearby or frequently co-accessed neighbor IDs are placed on the same page. 
However, maintaining such a layout in a dynamic index is difficult. As the topology evolves and new data are continuously inserted, static layouts quickly become outdated, while frequent global layout adjustments incur prohibitive maintenance overhead.

\etitle{(2) Reducing vector I/O without hurting accuracy.} Excessive vector I/O can be alleviated by separating approximate distance estimation from exact distance computation, \eg using compressed representations to filter candidates before reading full vectors. However, this alone is insufficient, because errors in approximate distances still require the system to read many full vectors to avoid missing true top-$k$ neighbors.


\stitle{DGAI}. To this end, we propose \oursys, an on-disk graph-based dynamic ANNS system that decouples the vector from graph topology, supporting efficient updates and high-performance query processing.
\oursys is built upon two techniques: 

First, to address the challenge of preserving topology locality under continuous updates, we propose a \textit{similarity-aware dynamic layout} mechanism that performs online layout optimization during insertion.
In graph-based ANNS, traversal iteratively expands toward nodes with higher similarity to the query. As a result, nodes that are close in the vector space tend to be explored within the search trajectory.
Therefore, placing such nodes on the same page can improve locality and reduce I/O.
Our key observation is that insertion already exposes similarity information by searching for similar existing nodes to identify candidate neighbors.
We therefore reuse this similarity information to guide placement, instead of following the append-only layout strategy commonly used by existing insertion methods. Specifically, a newly inserted node is placed into the page containing its most similar existing node. If the pages corresponding to the top few most similar nodes are all full, we split the page containing the closest node to create space for the insertion.

Second, to mitigate the additional random I/O incurred by raw-vector accesses under decoupled storage, we propose a \textit{two-stage query strategy enhanced by hierarchical product quantization}. 
In \oursys, graph traversal is first conducted using in-memory PQ-compressed vectors (without accessing on-disk vectors), followed by a final rerank stage using raw vectors.
This design avoids fetching raw vectors for the large number of intermediate candidates encountered during traversal. However, preserving high search accuracy still requires reranking many candidates with raw vectors, which introduces substantial random I/O.
By analysis, we find that the root cause is that PQ trained on global data produces "outlier" sub-vectors whose quantization leads to large distance distortions.
To address this limitation, we propose an optimization framework and instantiate it in PQ as hierarchical PQ.
The base layer models the overall data distribution, while the outlier layer specifically refines high-error regions.
This hierarchical structure substantially improves distance estimation, reducing redundant raw-vector accesses by 71\% with a modest 12\% increase in memory.

We compare \oursys with state-of-the-art dynamic ANNS systems, including FreshDiskANN~\cite{FreshDiskANN} and OdinANN~\cite{fast26odinann}.
Experimental results demonstrate that \oursys improves update speeds up to $8.1\times$ and query efficiency up to $2.5\times$ compared to both baselines. In mixed workloads involving concurrent insertions, deletions, and queries, \oursys achieves exceptional stability, cutting peak query latency by approximately 67\% compared to OdinANN and 63\% against FreshDiskANN.

In summary, we make the following contributions.

\begin{itemize}[leftmargin=0.2in]

\item[$\bullet$] 
We analyze the limitations of traditional coupled storage architecture and pinpoint the root causes of query performance degradation in decoupled designs (§\ref{motivation}).

\item[$\bullet$]
We propose \oursys, a dynamic on-disk ANNS system built upon an update-friendly decoupled storage architecture. It ensures low-latency query via two key techniques: similarity-aware dynamic layout and two-stage query strategy with hierarchical PQ  (§\ref{sec:insert}\&§\ref{sec:query}).

\item[$\bullet$]
We further introduce a suite of system-level optimizations to boost overall performance (§\ref{sec:implementation}). Extensive experiments on diverse datasets show that \oursys achieves significantly higher update throughput and lower query latency than state-of-the-art baselines under concurrent mixed workloads (§\ref{EXP}).

\end{itemize}

\eat{
\begin{itemize}[leftmargin=0.2in]

\item[$\bullet$] 
We propose a decoupled graph-based ANNS storage architecture that separates the vectors from the graph topology
based on a thorough analysis of existing dynamic ANNS systems. This decoupled design greatly improves the update performance of the on-disk graph-based index.

\item[$\bullet$] Building on this decoupled
architecture, we propose an efficient two-stage query strategy by leveraging the hierarchical PQs and a similarity-aware incremental topology reordering to minimize read amplification during graph traversal and result refinement.

\item[$\bullet$] We implement \oursys, a high-performance dynamic graph-based index for billion-scale ANNS tasks upon a decoupled storage architecture.
Extensive experiments on real-world datasets demonstrate that \oursys 
achieves substantial improvements in both update and query efficiency compared to state-of-the-art disk-based systems, while maintaining comparable accuracy.
\end{itemize}
}

\section{Background}



In this section, we introduce some fundamental concepts and key techniques of approximate nearest neighbor search and graph-based index updates.
\label{sec:graph-based-anns}

\subsection{Graph-based ANNS index}

\stitle{Approximate Nearest Neighbor Search (ANNS).} 
Given a dataset $\mathcal{V} = \{V_1, V_2, \ldots, V_N\} \subset \mathbb{R}^D$ of $N$ vectors in $D$-dimensional space and a query vector $V_q \in \mathbb{R}^D$, the ANNS aims to identify the $k$ vectors $\mathcal{V}^k_q \subset \mathcal{V}$
closest to the query $V_q$ based on some distance rules\footnote{Different graph-based index construction methods adopt different distance-based rules for neighbor pruning or filtering. This work is suitable for all the rules. 
}~\cite{DiskANN,HNSW,NSG,NSW}. 
The query accuracy is measured by $recall@k=(|\mathcal{V}^k_q \cap \mathcal{V'}^k_q|) /|\mathcal{V'}^k_q|$, where $\mathcal{V'}^k_q$ is the exact top-$k$ nearest neighbors of $V_q$ in $\mathcal{V}$. In this paper, we adopt recall@10 as the default accuracy metric.

\begin{figure}[t]
        
		\centering
		\subfloat[Vector data]{
            \label{fig:base_a}
		  \includegraphics[width=0.31\linewidth]{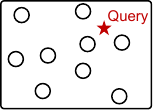}
        }
		\subfloat[Graph-based index]{
            \label{fig:base_b}
		  \includegraphics[width=0.31\linewidth]{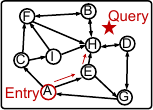}
        }
        \subfloat[Index layout]{
            \label{fig:base_d}
		  \includegraphics[width=0.31\linewidth]{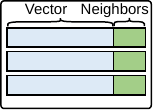}
        }
		\vspace{-0.05in}
        \caption{An example of a graph-based vector index on a vector dataset.}
        \label{fig:base_graph_index}
\end{figure}

\stitle{On-Disk Graph-Based Index.}
Graph-based index treats each vector as a node and constructs a directed graph based on distance rules. 
In Figure~\ref{fig:base_b}, we present an example of a graph-based index constructed for the vector dataset 
in Figure~\ref{fig:base_a}. For each node, the outdegree is bounded by a predefined threshold $R$, thereby preventing query performance degradation due to the overhead of calculating and comparing distances with an excessive number of neighbors during ANNS. For scalability to billion-scale datasets, modern graph-based ANNS systems typically store their indices on disk~\cite{osdi25pipeann,FreshDiskANN,DiskANN,IP-DiskANN,pgvector,fast26odinann}.
They all adopt a coupled layout, where each vector and its corresponding neighbor list are stored contiguously within the same disk page (Figure~\ref{fig:base_d}).

\begin{figure}[t]
    \centering
    \includegraphics[width=1\linewidth]{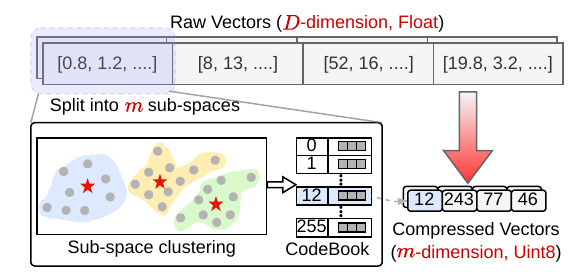}
    \vspace{-0.1in}
    \caption{
       Illustration of product quantization.
    }
    \label{fig:pq}
    \vspace{-0.08in}
\end{figure}

\stitle{Product Quantization.} 
To reduce costly vector accesses, most on-disk ANNS systems~\cite{DiskANN,osdi25pipeann,fast26odinann,FusionANNS,Starling,spfresh,SPANN,Greator,FreshDiskANN} adopt product quantization (PQ) to maintain a compact in-memory representation of vectors for low-cost distance estimation during traversal.
As shown in Figure~\ref{fig:pq}, PQ decomposes each $D$-dimensional vector into $m$ low-dimensional subspaces, \ie $m$ sub-vectors, and performs $K$-means (\eg $K=256$) clustering independently within each subspace to learn a codebook (\ie the $K$ centroids).
For each vector, its sub-vectors are approximately represented by their corresponding centroids. Each sub-vector is then encoded by the ID of its assigned centroid rather than the centroid itself, enabling the raw vector to be compressed into $m$ dimensions.
Recent studies further extend PQ to dynamic scenarios~\cite{laq,fast26odinann,Greator,FreshDiskANN}, allowing compact representations to be updated efficiently under insertions and deletions.
However, because PQ is a lossy compression, it inevitably discards fine-grained information within each subspace, introducing a non-negligible approximate error.

\newcommand{\argmin}{\mathop{\mathrm{arg\,min}}}
\begin{algorithm}[t]
\SetAlgoNoLine
\SetAlgoSkip{0pt}
\SetAlCapSkip{0pt}
\SetInd{0.3em}{0.6em}
\setlength{\interspacetitleruled}{0.5ex}
\setlength{\interspacealgoruled}{0.5ex}
\caption{Best-first Greedy Search\cite{azizi2025graph}}
\label{alg:pq_search}
\KwIn{
    Graph $G(V, E)$, query vector $V_q$, entry node $V_e$, \\
    result size $k$, candidate queue size $l \geq k$
}
\KwOut{$k$ approximate nearest neighbors to $V_q$}

    $C \gets \{(V_e, \mathrm{dist}_{\mathrm{PQ}}(V_q, V_e))\}$\tcp*[r]{(id, PQ dist)}
    $Q \gets \emptyset$\tcp*[r]{(id, Exact dist)}
    
    \While{$C \setminus V(Q) \neq \emptyset$}{\label{start_stage1}
        $(p^*, \delta_{\mathrm{pq}}) \gets \argmin_{(u, \delta) \in C \setminus V(Q)} \delta$\;\label{alg:compute}
        
        Load vector $V_{p^*}$ and $N_{\mathrm{out}}(p^*)$ from disk\;\label{alg:load_nei}\label{greedy:line:5}
        
        $\delta_{\mathrm{exact}} \gets \mathrm{dist}_{\mathrm{exact}}(V_q, V_{p^*})$\tcp*[r]{Exact dist}

        $Q \gets Q \cup \{(p^*, \delta_{\mathrm{exact}})\}$\;
        
        \ForEach{$u \in N_{\mathrm{out}}(p^*)$}{
            $C \gets C \cup \{(u, \mathrm{dist}_{\mathrm{PQ}}(V_q, u))\}$\tcp*[r]{PQ dist}
        }
        
        \If{$|C| > l$}{
            $C \gets \text{Top-}l\text{ closest nodes in } C$\;
        }
    }\label{end_stage1}
    
    \Return the $k$ nodes in $Q$ with smallest exact distances\;

\end{algorithm}

\stitle{Best-first Greedy Search.}
In ANNS, graph-based methods utilize a best-first greedy search\cite{HNSW,DiskANN,osdi25pipeann,azizi2025graph}.
As shown in Algorithm~\ref{alg:pq_search}, best-first greedy search maintains an ANN candidate queue with a fixed size, containing the current top-$l$ nearest vectors sorted by distance. 
The search starts with a fixed entry node and gets closer to the query vector step by step. Each step will expand the current closest node and add its neighbors to the candidate queue, until all nodes in the queue are expanded.
This greedy graph traversal quickly retrieves approximate nearest neighbors without scanning the entire dataset.

In existing systems, approximate evaluation and exact distance computation are tightly integrated during search.
When a node is accessed, its topology and raw vector are fetched together, and the system performs exact distance evaluation for the current node while using approximate distances to assess its neighbors. This execution model is enabled by the coupled layout of topology and vectors, which improves I/O efficiency by requiring only a single data fetch.


\eat{
\newcommand{\argmin}{\mathop{\mathrm{arg\,min}}}
\begin{algorithm}[t]
\small
\SetAlgoNoLine
\SetAlgoSkip{0pt}
\SetAlCapSkip{0pt}
\SetInd{0.3em}{0.6em}   
\setlength{\interspacetitleruled}{0.5ex} 
\setlength{\interspacealgoruled}{0.5ex}  
\caption{Best-first Greedy Search\cite{azizi2025graph}}
\label{alg:pq_search}
\KwIn{
    Graph $G(V, E)$, query vector $V_q$, entry node $V_e$, \\
    result size $k$, candidate queue size $l \geq k$
}
\KwOut{$k$ approximate nearest neighbors to $V_q$}

    $C \gets V_e$\tcp*[r]{ANN candidate queue}
    $Q \gets \emptyset$\tcp*[r]{visited set}
    \While{$C \setminus Q \neq \emptyset$}{\label{start_stage1}
        $p^* \gets \argmin_{V_i \in C \setminus Q} \mathrm{dist\_function}(V_q, V_i)$\;\label{alg:compute}
        Load $N_{\mathrm{out}}(p^*)$ from disk\;\label{alg:load_nei}
        $C \gets C \cup N_{\mathrm{out}}(p^*)$\;
        $Q \gets Q \cup \{p^*\}$\;
        \If{$|C| > l$}{
            $C \gets \text{Top-}l\text{ closest nodes in } C \text{ to } V_q$\;
        }
    }\label{end_stage1}
    \Return the $k$ closest nodes in $C$\;

\end{algorithm}
}



\subsection{Index Updates in Vector Search}
\label{sec:updates_background}

In real-world applications, vector databases must support dynamic data updates (insertions and deletions) to dynamically maintain a high-quality index and ensure efficient query performance.

\textbf{Insertions.} Logically, the insertion operation consists of two steps. The first step performs a similarity search for the newly inserted vector to find its candidate neighbors within the database. The second step involves adding edges: in addition to creating edges from the new node to its neighbors, the system must also establish edges from existing nodes directed to the new node to facilitate efficient routing in subsequent graph traversals.

\textbf{Deletions.} To prevent the degradation of graph index quality caused by deleted nodes, mainstream methods typically employ a \textit{full link repair} strategy\cite{FreshDiskANN,fast26odinann}. Specifically, they reconnect all the in-neighbors and out-neighbors of the deleted node to patch the search paths broken by the deletion. However, since the in-neighbors of a deleted node cannot be directly accessed in a standard directed graph, this repair step necessitates extensive index traversals, triggering a massive number of disk I/O requests. To alleviate this issue, existing solutions generally adopt a \textit{lazy delete} mechanism combined with batch processing to amortize the substantial I/O costs.
\section{Motivation}
\label{motivation}


In this section, we first analyze existing graph-based ANNS systems that couple the topology with vectors and point out their limitations in index updates.
We then investigate the root causes of query performance degradation in the decoupled design.

\subsection{Limitations of Updates in Coupled Index Storage} 
\label{sec:motivation:limit}

As discussed in the Introduction, the coupled architecture is unfriendly to updates. In this subsection, we will provide a detailed analysis and motivate the purpose of the decoupled storage architecture.

\begin{figure}[tbp]
	\centering
	\begin{minipage}[t]{0.48\linewidth}
		\centering
		\includegraphics[width=\linewidth]{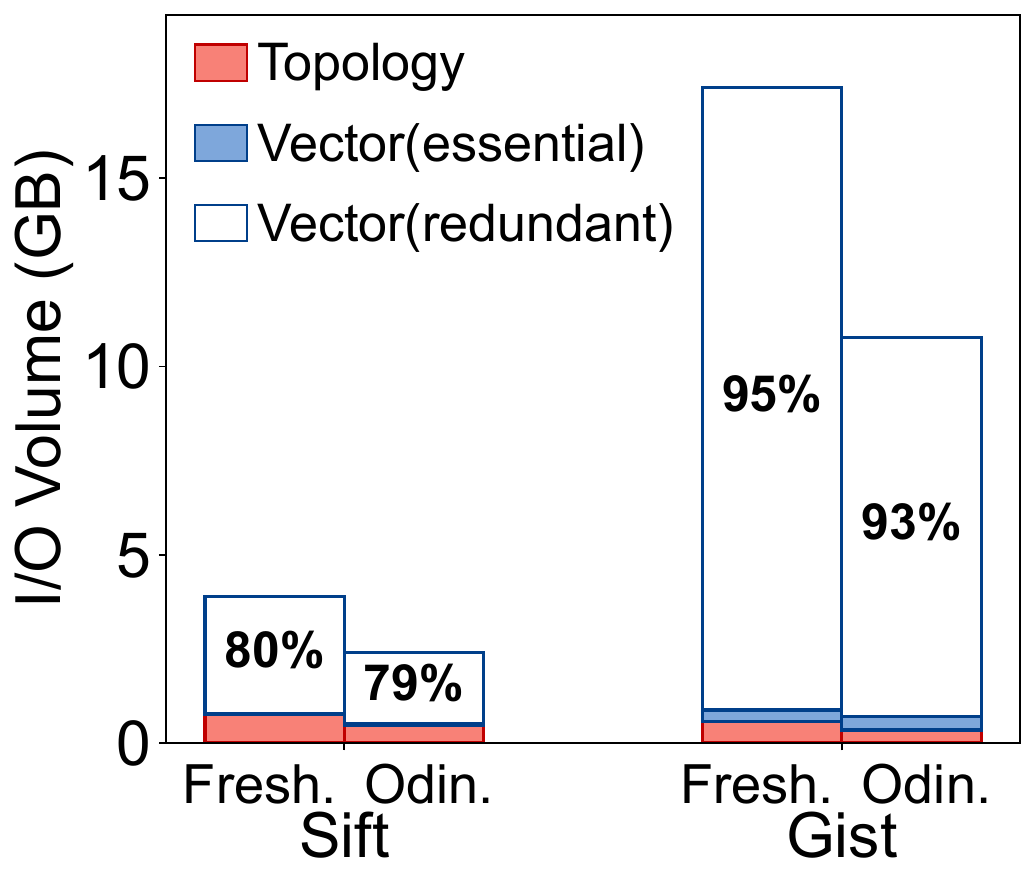}
		\vspace{-0.2in}
        \caption{I/O volume breakdown during index update.}
		\label{fig:effectiveio}
	\end{minipage}
	\hfill 
	\begin{minipage}[t]{0.49\linewidth}
		\centering
 		\includegraphics[width=\linewidth]{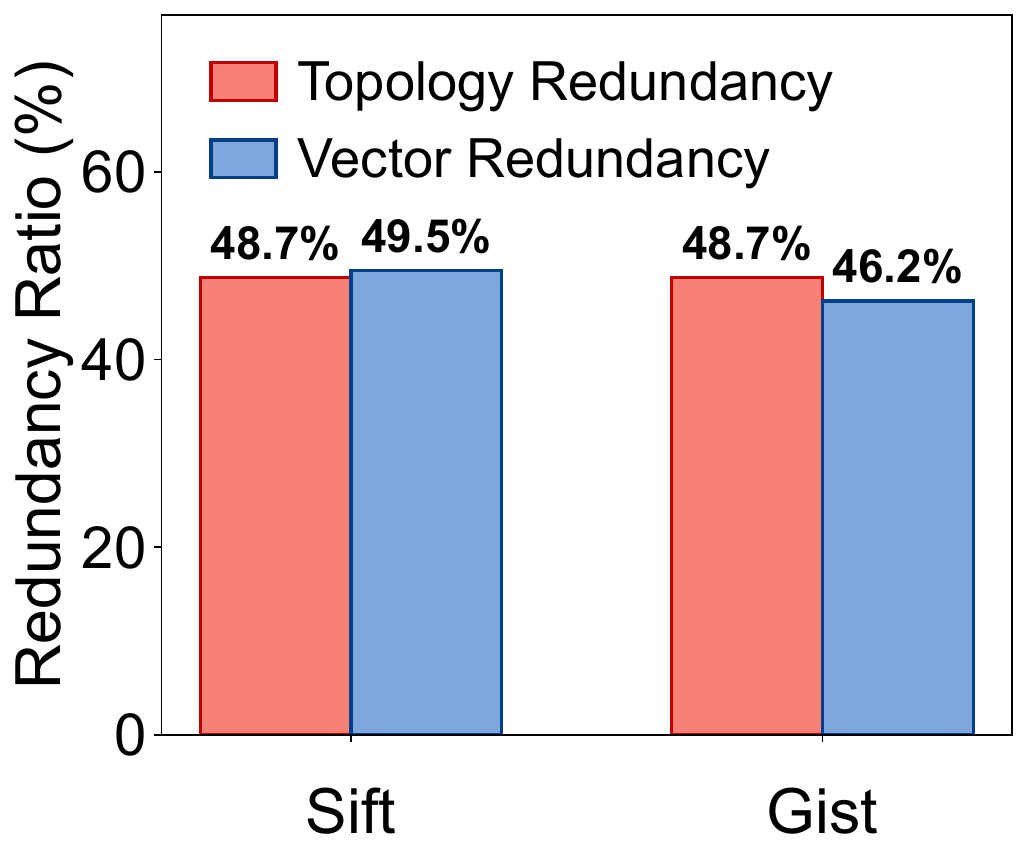}
		\vspace{-0.2in}
        \caption{I/O redundancy breakdown during query.}
		\label{fig:queryImprovement}
	\end{minipage}
    \vspace{-0.05in}
\end{figure}

\stitle{I/O Breakdown and Analysis.}
To identify the source of the I/O bottleneck (as shown in Figure~\ref{fig:update_composition}), we analyze the I/O composition of two representative systems (FreshDiskANN~\cite{FreshDiskANN} and OdinANN~\cite{fast26odinann,OdinANNgithub}) during index updates.
Figure~\ref{fig:effectiveio} presents the breakdown of I/O volume for topology and vector reads.
While vector I/O dominates the total volume, we found that a large fraction is redundant, with many invalid vector reads.

As shown in \eat{the }Figure~\ref{fig:effectiveio}, our analysis reveals that this redundancy constitutes more than \red{79\%} of all I/O.
This massive inefficiency stems from the tightly coupled storage architecture common in existing graph-based ANNS systems, where each node’s vector and topology information are stored together on the same disk page.
This coupled design forces the system to load a node’s vector and topology information together even when only topology is needed, for example, when updating the topology of tens to hundreds of neighbors affected by the updated node during the update process.
\eat{Because}Given that vector data is significantly larger than topology data (\eg on the Gist dataset with a threshold of 32 neighbors, vector data accounts for 97\% of the total index space), \eat{significant}a substantial amount of I/O bandwidth is used to read the invalid vector data. 
Based on this observation, a direct approach to resolve this inefficiency is to decouple the storage of topology and vectors, thereby ensuring that topology-only operations do not incur accessing redundant vector data.

\etitle{I/O Redundancy during Update}. The significant redundant I/O overhead caused by the coupled storage architecture during updates motivates us to decouple the topology from the raw vector, enabling index updates to be performed efficiently.

\subsection{Limitations of Query in Decoupled Index Storage}
\label{sec:motivation:challenge}

Although the decoupled storage architecture improves update performance, we observe that it leads to a degradation in query performance, as shown in Figure \ref{fig:query_composition} in the Introduction. 
This subsection analyzes the root causes of this degradation, identifying two distinct forms of I/O problems derived from decoupled architecture.

\eat{
Although the decoupled storage architecture improves update performance, we observe that it degrades query performance, as shown in Figure~\ref{fig:query_composition}. 
To mitigate the I/O bottleneck discussed in Section~\ref{sec:motivation:challenge}, a straightforward solution is to separate graph traversal from exact distance computation. Then the query is divided into two stages.
\textit{In the first stage}, the greedy search traverses the graph using only PQ-derived approximate distances, thereby constructing the nearest-neighbor (NN) candidate queue without accessing original vectors.
\textit{In the second stage}, the original vectors of the top-$\tau$ candidates in the queue are retrieved, and their exact distances to the query vector are computed to produce an accurate top-$k$ nearest neighbors.

By avoiding exact vector reads for all nodes along the search path and enabling batched vector I/O in the reranking stage, this design improves query performance by \red{10\%--14\%} over the purely decoupled baseline on \red{Sift and Gist} (Figure~\ref{fig:queryImprovement}). 
\textit{However, it still lags behind the coupled design by \red{13\%--17\%} under the same recall target}. 
In the following, we analyze the root causes of this residual overhead and show that it stems from two distinct forms of I/O inefficiency.
}

\stitle{Topology Read Amplification.}
\label{sec:motivation_reorder}
In the decoupled architecture, separating vectors from the topology implies that reading a single node's topology information inevitably retrieves a substantial amount of data regarding other nodes. This stems from the storage system's mandatory 4KB fetch unit. In the coupled architecture, heavyweight vectors dominated page capacity, naturally minimizing the retrieval of invalid data. In contrast, the decoupled design suffers from severe read amplification, where valid topology information constitutes only 3\% of a fetched page (with a maximum degree of 32).

\stitle{Excessive Vector I/O.}
In the coupled architecture, the overhead of vector access is naturally hidden, as vectors and topology are co-located, allowing a single I/O operation to retrieve both during node expansion. After decoupling, however, vector retrieval becomes an independent source of I/O overhead.
Moreover, during graph traversal, many expanded nodes act mainly for routing and typically do not appear in the final top-$k$ results. Consequently, fetching vectors for these intermediate candidates results in a significant volume of unnecessary I/O, which persists as a primary bottleneck for query efficiency in decoupled designs.

\etitle{I/O Redundancy during Query.}
Overall, in the decoupled architecture both the reading of topology pages and the reading of vector pages exhibit significant redundancy (Figure~\ref{fig:queryImprovement}).
To tackle these two primary I/O bottlenecks, we propose two core designs in our system.
(§\ref{sec:insert}\&§\ref{sec:query}).

\section{Overview}
\label{sec:querydesign}

\begin{figure}[t]
    \centering
    \includegraphics[width=\linewidth]{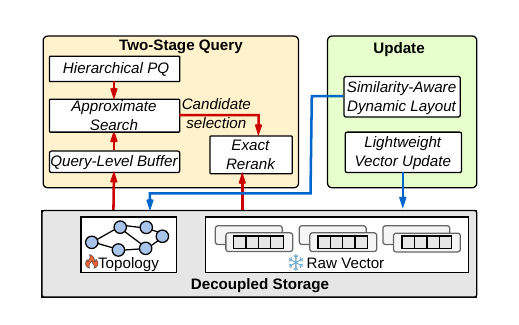}
    \vspace{-0.3in}
    \caption{
        Workflow of the \oursys.
    }
    \label{fig:arch}
    \vspace{-0.05in}
\end{figure}

Based on the above motivation and observation, we propose a new \underline{d}ecoupled on-disk \underline{g}raph-based \underline{A}NNS \underline{i}ndexing system, \oursys.
In this section, we present the design of \oursys from three aspects: storage, query and update. 

\eat{
To begin with, we introduce \oursys's workflow in the query and update process (Section~\ref{sec:workflow}).
We then
introduce a two-stage query based on hierarchical PQ (Section~\ref{sec:query}),
which filters out vast redundant overhead and effectively resolves the additional I/O issue during query brought by the decoupled design. Finally, we propose an incremental page-level reordering strategy based on vector similarity (Section~\ref{sec:insert}),
which places similar nodes on the same page. By coordinating with the caching strategy, this approach effectively leverages fetched redundant data and mitigates the read amplification problem.
}

\label{sec:workflow}


\stitle{Decoupled Storage.}
As illustrated in Figure~\ref{fig:arch}, \oursys decouples graph topology from raw vectors by storing them in separate on-disk files. To facilitate low-cost distance estimation during search, the system maintains compressed vector representations in memory. This decoupled storage architecture improves I/O efficiency during updates by avoiding the overhead of rewriting large raw vector for topology-only modifications, and provides the basis for topology layout optimizations introduced later.

\stitle{Hierarchical PQ Enhanced Two-Stage Query.}
As illustrated in the left part of Figure~\ref{fig:arch} (highlighted in red), the query process employs a hierarchical PQ enhanced 
two-stage query strategy, augmented by a query-level buffer to accelerate graph traversal.
\oursys first employs a \textbf{two-stage query} scheme:
\textit{In the first stage}, the greedy search traverses the graph using only PQ-derived approximate distances, thereby constructing the nearest-neighbor (NN) candidate queue without accessing raw vectors.
\textit{In the second stage}, the raw vectors of the top-$\tau$ candidates in the queue are retrieved, and their exact distances to the query vector are computed to produce an accurate top-$k$ nearest neighbors.
To minimize the vectors needed to be fetched in the second stage, \oursys employs \textbf{hierarchical PQ}. 
In the hierarchical PQ framework, two types of codebooks are employed: a base codebook constructed from the entire dataset, and outlier codebooks built from subsets of data. The outlier codebooks focus on vectors for which the base codebook incurs large quantization errors, thereby enhancing the accuracy of candidate ranking. To reduce computation and I/O overhead, only the top-$\tau$ ($\tau \ge k$) nodes in the candidate queue are selected for exact distance evaluation, with their raw vectors fetched from disk. The final top-$k$ nearest neighbors are then derived from these precise distance computations, minimizing unnecessary disk access and redundant calculations.

\stitle{I/O Efficient Update Enhanced by Decoupled Storage.}
The update process is illustrated by the blue arrows in Figure~\ref{fig:arch}.
\textbf{For insertions}, to better leverage the decoupled storage architecture, we propose a similarity-aware dynamic layout that is maintained during insertions. By colocating nodes that are likely to be accessed in close succession during search, this design improves page reuse, and significantly lowers the number of disk page writes during insertion.
Meanwhile, the raw vector side only requires a lightweight vector update, \ie writing the new vector into an available slot.
\textbf{For deletions}, \oursys achieves high I/O efficiency thanks to the separation of topology and vector storage. To remove a node, the system simply marks the obsolete vector as deleted and updates the graph topology using existing methods \cite{liuwolverine,pgvector,FreshDiskANN}. Crucially, this completely eliminates the need to fetch raw high-dimensional vectors from disk, thereby saving massive I/O bandwidth.

\eat{
The update process is illustrated by the blue arrows in Figure~\ref{fig:arch}. Thanks to the decoupled architecture of topology and vectors, updates in \oursys are divided into two independent procedures: topology updates and vector updates. 
This separation avoids loading irrelevant data during each procedure, thereby reducing I/O overhead and improving update performance.
For topology updates, though \oursys updates the index topology by employing the existing index update methods \cite{liuwolverine,pgvector,FreshDiskANN}, the flexibility of decoupled design allows it to update efficiently without redundant I/O caused by vectors.
For vector updates, \oursys either writes the new vector into the disk or marks the obsolete vector as deleted.
To better leverage the decoupled storage architecture, we propose an incremental page-level similarity-aware reordering method for the topology, which enhances data locality and improves cache utilization during update and query processes.
}

\section{
Dynamic and Prefetchable Similarity-Aware Topology Storage
}
\label{sec:insert}
\label{sec:design:reorder_layout}

To address inefficiency of topology access, we propose a similarity-aware dynamic layout maintained by insertion. Rather than treating insertion only as a graph maintenance operation, \oursys uses it as an opportunity to continuously optimize the topology layout. 

\subsection{Prefetching Opportunity Enabled by Decoupled Storage}\label{locality}






As discussed in Section~\ref{sec:motivation:challenge}, under the decoupled layout, topology pages exclude the large raw vectors, allowing a single page to hold topology information for many more nodes. However, each expansion step in greedy search typically requires only the neighbor IDs of one node, while disk I/O must still be issued at the 4KB page granularity. This mismatch causes each read to fetch a large amount of data irrelevant to the current expansion, leading to severe read amplification; in practice, only about 3\% of the fetched data is useful.

\stitle{Opportunity.}
Fortunately, the extra data brought in by each read are still topology data and may be useful in subsequent search steps. When such reuse occurs, later accesses can be served from pages already resident in memory, thereby avoiding additional expensive disk I/O. In this way, the read amplification can be turned into useful prefetching rather than wasted overhead.
\eat{
This read amplification also opens up a new opportunity in decoupled storage. In a coupled layout, much of each page is occupied by raw vectors. Although necessary for exact distance computation, these vectors are irrelevant to topology expansion during graph traversal and thus limit how much topology information a page can hold. \gsf{comment}
By contrast, decoupled topology pages exclude raw vectors, allowing each page to cover the topology of many more nodes.
This creates opportunities for prefetching: if the topology data brought in by one page read is effectively utilized, subsequent node accesses are more likely to hit pages already in memory, thereby turning part of the amplified reads into useful prefetches.}

To realize this opportunity, nodes that are likely to be accessed in close succession should be co-located on the same pages.
However, maintaining such locality is difficult for two reasons.
First, in dynamic systems, frequent insertions and deletions continuously change the graph topology. Therefore, an effective layout optimization strategy should remain robust to topology evolution.
Second, even with a robust strategy, repeatedly applying static global reordering is not a practical way to maintain locality, as it incurs prohibitive computational overhead in real-time dynamic systems.
We therefore need a dynamic layout strategy that is both robust to topology evolution and efficient to maintain online. To guide the design of such a strategy, we next analyze the search trajectory of best-first greedy search.

\subsection{Dynamic Similarity-Aware Topology Layout}
\label{sec:reorder}

\eat{
Based on the above analysis, we design an incremental topology reordering method guided by vector similarity and tightly integrated with insertion. The core principle is simple: since graph traversal tends to access similar nodes together, the physical layout should place such nodes in the same page whenever possible. By enforcing this principle during insertion, \oursys continuously preserves locality as the index evolves, avoiding the need for expensive global reordering.
}

\begin{figure}[t]
    \centering
    \includegraphics[width=0.85\linewidth]{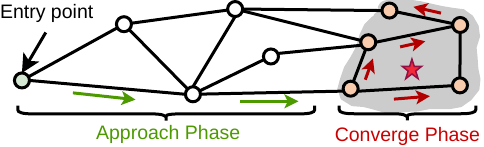}
    \vspace{-0.05in}
    \caption{Two phases of best-first greedy search.
    }
    \label{fig:tp}
    \vspace{-0.15in}
\end{figure}

As shown in Figure~\ref{fig:tp}, best-first greedy search can be logically viewed as consisting of two phases: approach and converge~\cite{osdi25pipeann}. In the approach phase, the search quickly moves toward the query vector. In the converge phase, which dominates the overall search time\cite{zhou2025govector, PANNS}, 
the search performs gradual expansions within a small space region to identify the top-$k$ nearest nodes. 
As a result, many of the accessed nodes during the converge phase are similar to one another.

This locality implies that if similar nodes are co-located on the same page, data fetched for one expansion can naturally serve as useful prefetching for subsequent expansions. In a dynamic index, the key challenge is how to maintain such similarity-aware co-location efficiently under continuous updates.

\stitle{Page-Level Maintenance on Insertion.}
Insertion is well suited for preserving locality in a dynamic index, because it involves a nearest neighbor search procedure for the new node, which provides the similarity information needed for layout optimization.
Since online maintenance must remain efficient and disk I/O occurs at page granularity, \oursys enforces locality at the page level: instead of appending the new node to an arbitrary free position\cite{fast26odinann,FreshDiskANN}, 
it places the node onto the page containing its nearest existing neighbor. In this way, \oursys preserves locality for incremental updates with little additional overhead by leveraging similarity information already computed during insertion.

\stitle{\textsc{SADL}.}
Based on the above analysis, we design \underline{s}imilarity-\underline{a}ware 
\underline{d}ynamic \underline{l}ayout (\textsc{SADL}), a dynamic layout mechanism that maintains physical locality during insertion by placing each newly inserted node close to its most similar neighbors. As outlined in Algorithm~\ref{alg:insert}, \textsc{SADL} consists of three phases:

\textit{(1) Similarity-Driven Pages Selection (line \ref{alg:line1}-\ref{alg:line2}):}  
When inserting a new node $v$, the standard insertion procedure already performs a greedy search to find its nearest neighbors within the existing data. We leverage this search result to identify the target page—specifically, the disk page containing the node $u$ that is most similar to $v$. This ensures that our placement strategy is aligned with the access locality exhibited during query processing.
\textit{(2) In-Place Insertion Attempt (line \ref{alg:line3}-\ref{alg:line6}):}  
The system first attempts to insert $v$ directly into the target page. If the page has sufficient free slots, the insertion completes with a single I/O write, instantly preserving perfect locality.
\textit{(3) Similarity-Based Page Splitting (line \ref{alg:line7}-\ref{alg:line21}):}  
If the target page is at capacity, we trigger a page split operation governed by vector similarity.
Half of the nodes in this page are repartitioned into a new page according to the similarity relation between all nodes in the page.
Finally, the new node is inserted into the page where the nearest node resides.

\begin{algorithm}[t]
\SetAlgoNoLine
\SetAlgoSkip{0pt}
\SetAlCapSkip{0pt}
\SetInd{0.3em}{0.6em}   
\setlength{\interspacetitleruled}{0.5ex} 
\setlength{\interspacealgoruled}{0.5ex}  
\caption{\textsc{SADL}}
\label{alg:insert}
\KwIn{New vector $v$}
\KwOut{Target page of $v$}

$N \gets GreedySearch(v)$\tcp*[r]{Sorted in ascending order of distance to $v$}\label{alg:line1}
$\mathcal{P} \gets \{ \texttt{GetPage}(u) \mid u \in N \}$\; 
\label{alg:line2}
\ForEach{$p \in \mathcal{P}$}{\label{alg:line3}
    \If{$p.size() + |Nout(v)| \le MaxCapacity$}{
        Insert $v$ into page $p$\;
        \Return $ p $ \;\label{alg:line6}
    }
}
$p_{old} \gets \texttt{Getpage}(N[0]) $\;\label{alg:line7}
$ S \gets \texttt{GetNodes}(p_{old}) $\tcp*[r]{Collect node IDs and neighbors from the page to be split.}
$\texttt{ClearPage}(p_{old}) $\;
$p_{new} \gets \texttt{NewPage}()$\;
\ForEach{$ (u, Nbrlist) \in S $}{
    \If{$ u \notin p_{old} \cup p_{new} $}{
        $p_{\text{target}} \gets \arg\min_{p' \in \{p_{old}, p_{new}\}} p'.size()$\;
        Insert $ u $ into $ p_{target}$\;
    }
    \Else{
        $p_{\text{target}} \gets \texttt{GetPage}(u)$\;
    }
    \ForEach{$ w \in (Nbrlist \cap S) \text{ and } w \notin (p_{old} \cup p_{new}) $}{
        \If{$ p_{target}.size() + |Nout(w)| \le MaxCapacity/2$}{
            Insert $ w $ into $ p_{target}$\;
        }
    }
}   
$ p^* \gets \texttt{GetPage}(N[0])$\;
Insert $v$ into $p^*$\;\label{alg:line21}
\Return $ p^* $\;

\end{algorithm}

\stitle{Overhead and Benefit Analysis.}
\textsc{SADL} adds only a lightweight placement cost to the original insertion procedure. Specifically, it checks the pages corresponding to the candidate neighbors and performs a simple placement decision, incurring an $O(R)$ overhead, 
where $R$ denotes the number of retrieved neighbors. This cost is negligible compared with the dominant cost of insertion, namely the neighbor search phase, whose complexity is $O(\log N)$. 
More importantly, \textsc{SADL} improves the on-disk layout of graph topology, thereby increasing page reuse during both the search phase and the subsequent reverse-edge maintenance phase of insertion. Consequently, it can substantially improve overall insertion performance. We validate these benefits experimentally in Section~\ref{exp:update_}.

\section{HPQ Enhanced Two-Stage Query}
\label{sec:query}
\eat{
In this section, we first present a two-stage query scheme tailored for decoupled topology–vector architectures. We then introduce a hierarchical PQ mechanism designed for this two-stage query workflow to further reduce both I/O overhead and computational cost.}
To reduce the overhead of fetching raw vectors, we propose a hierarchical PQ enhanced two-stage query strategy, allowing \oursys to achieve the target recall while fetching significantly fewer raw vectors from disk.

\begin{figure}[tbp]
	\centering
	\begin{minipage}[t]{0.48\linewidth}
		\centering
		\includegraphics[width=\linewidth]{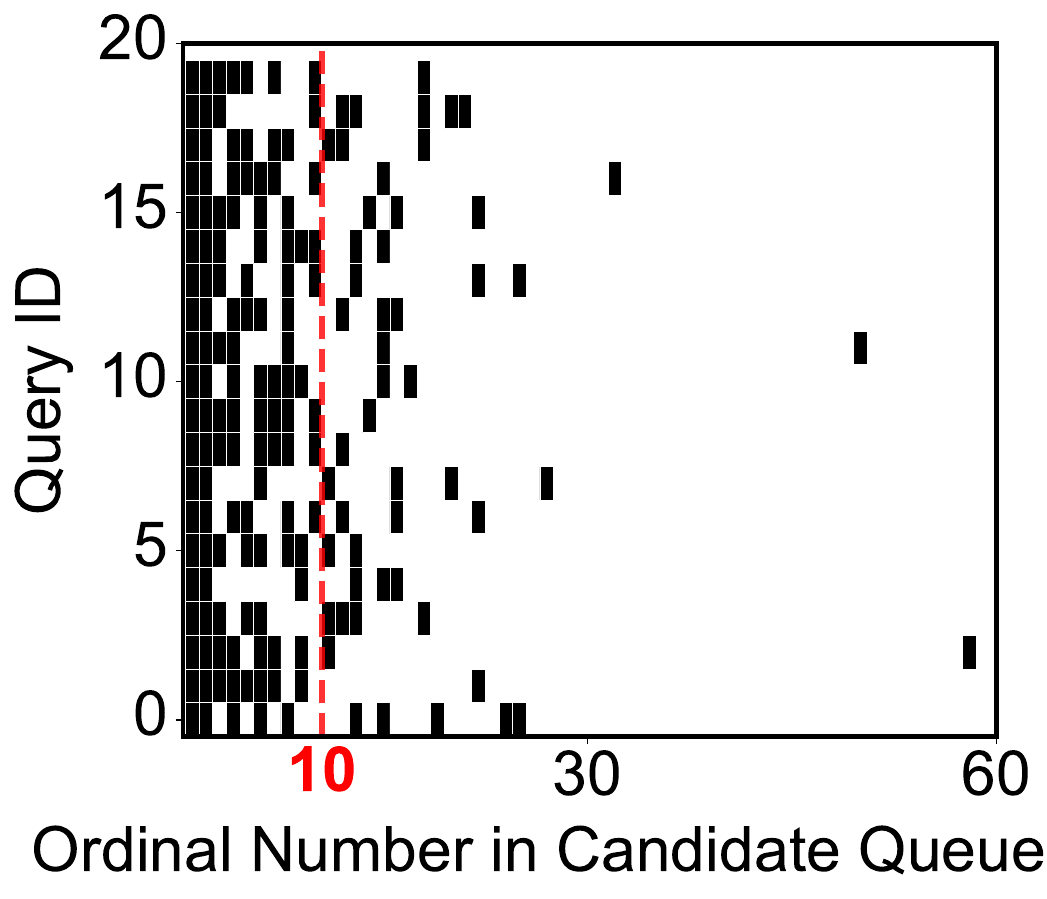}
        \vspace{-0.15in}
		\caption{The distribution of true NNs in the candidate queue ($k=10$).}
		\label{fig:distribution_new}
	\end{minipage}
	\hfill 
	\begin{minipage}[t]{0.49\linewidth}
		\centering
		\includegraphics[width=\linewidth]{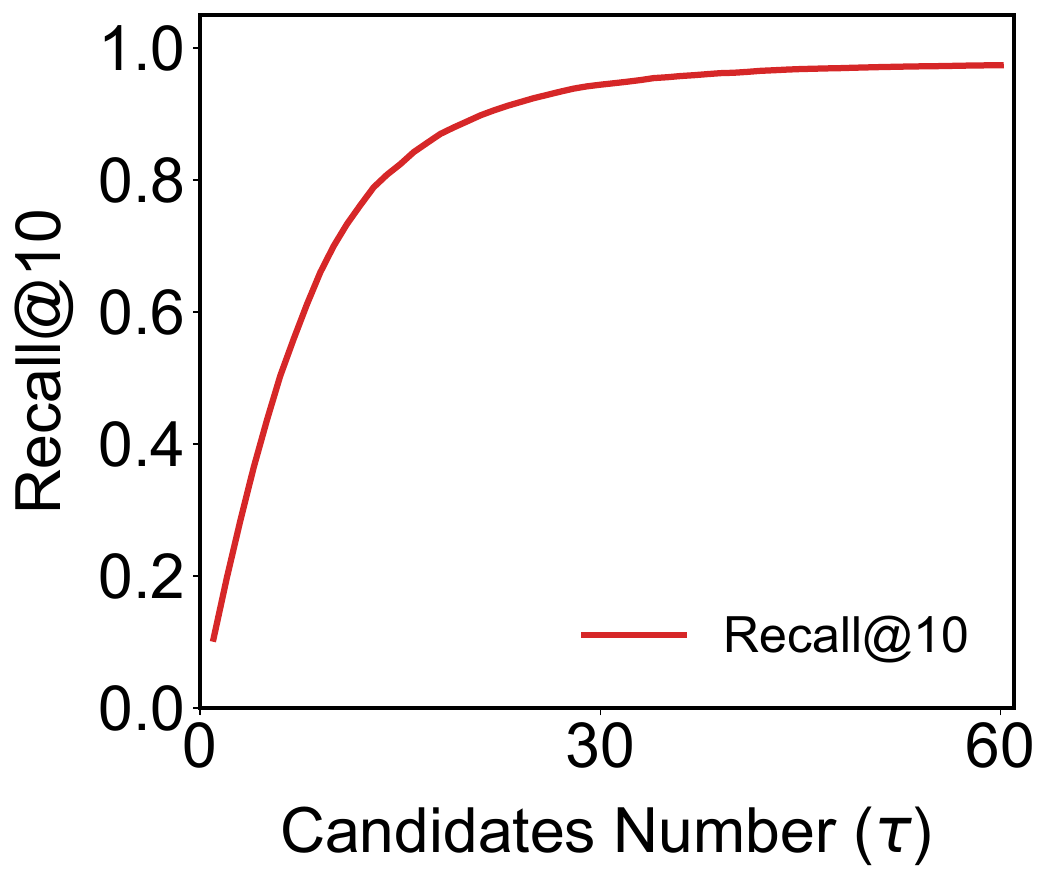}
        \vspace{-0.15in}
		\caption{Recall improves as the number of candidates ($\t$) increases.}
		\label{fig:recall_candidate}
	\end{minipage}
\end{figure}


\subsection{Naive Two-Stage Query}
To mitigate the excessive vector I/O during reranking discussed in Section~\ref{sec:motivation:challenge}, a straightforward solution is to separate approximate graph traversal from exact distance computation: the query first performs greedy search using only PQ-derived approximate distances to construct the nearest-neighbor (NN) candidate queue without accessing raw vectors, and then retrieves the raw vectors of the top-$\tau$ candidates to compute their exact distances and produce the top-$k$ nearest neighbors.
Although this method avoids raw vector reads for intermediate nodes during graph traversal and improves over the decoupled baseline, it still cannot match the performance of the coupled system at the same recall.

\stitle{Limitation.}
The primary bottleneck of the two-stage query is the need to retrieve more raw vectors due to PQ errors. As shown in Figure~\ref{fig:distribution_new}, while most true NNs are located near the head of the candidate queue, a critical fraction is displaced to the tail of the candidate queue. Consequently, a large $\tau$ is required to fetch them.
Although a larger $\tau$ can yield higher recall, the recall improvement diminishes rapidly as $\tau$ increases, as shown in Figure \ref{fig:recall_candidate}. This is because true NNs become sparse near the tail of the candidate queue, leading to significant invalid I/O overhead in the second stage.
\eat{
A recent study \cite{FusionANNS} adopts a periodic evaluation mechanism to dynamically determine the termination threshold. While this approach can reduce the total I/O volume, we observe that it splits a single batched vector fetch into multiple fragmented I/O requests. Consequently, this inadvertently increases the overall search latency, exerting a negative impact on system performance.
}

\begin{figure}[t]
    \centering
    \includegraphics[width=\linewidth]{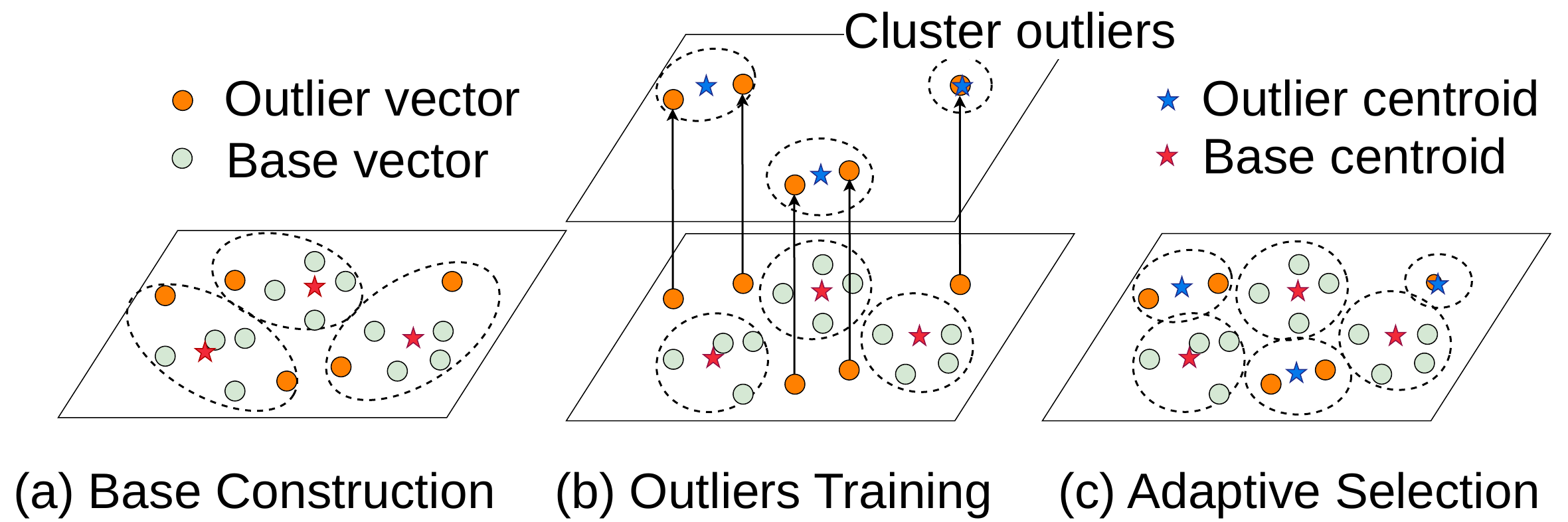}
    \vspace{-0.2in}
    \caption{Hierarchical PQ Construction Process.
    }
    \label{fig:hpq}
    \vspace{-0.1in}
\end{figure}

\subsection{Hierarchical PQ} 

\eat{
While PQ works well for vectors close to their centroids, it produces large distance errors for "outliers" that lie significantly far from their centers. 
}
Standard PQ decomposes a vector into $M$ sub-vectors and maps each to the nearest centroid in a codebook. While PQ works well for sub-vectors close to their centroids, sub-vectors located far from their assigned cluster centers suffer from severe quantization errors (\ie the L2 distance between the sub-vector and its assigned centroid).
Although increasing the number of centroids improves average quantization quality, it tends to make easy cases more accurate rather than resolving the hard cases that dominate the error tail.
As a result, a small fraction of poorly quantized sub-vectors can still disproportionately degrade the overall distance estimation, pushing true NNs to the tail of the candidate queue and forcing the system to fetch a massive $\tau$ with expensive disk I/Os.

\stitle{Hierarchical PQ.}
This observation reflects a more fundamental issue beyond PQ: rare but high-error cases are not well handled by a single globally trained quantizer. To address this, we propose a hierarchical compensation framework, instantiated here as hierarchical PQ.
As shown in Figure~\ref{fig:hpq}, HPQ consists of three stages:
(1) In \textit{Base Construction}, a base codebook $C_{base}$ is learned over all vectors via standard K-means to capture the overall data distribution. This codebook serves as the default quantizer and provides a unified approximation for the majority of sub-vectors.
(2) In \textit{Outliers Training}, for each subspace, we evaluate the quantization errors of all training sub-vectors under $C_{base}$ and select the top $\alpha\%$ (\eg $\alpha = 20$) with the largest errors. A dedicated outlier codebook $C_{outlier}$ is then trained on this subset. In this way, the additional modeling capacity is focused on the cases that contribute most to extreme quantization errors.
(3) In \textit{Adaptive Selection}, each sub-vector is encoded by either $C_{base}$ or $C_{outlier}$, depending on which yields the lower pq error. This adaptive routing preserves the efficiency of the base codebook for regular cases while allowing difficult cases to benefit from specialized modeling.

\begin{figure}[t]
    \centering
    \includegraphics[width=\linewidth]{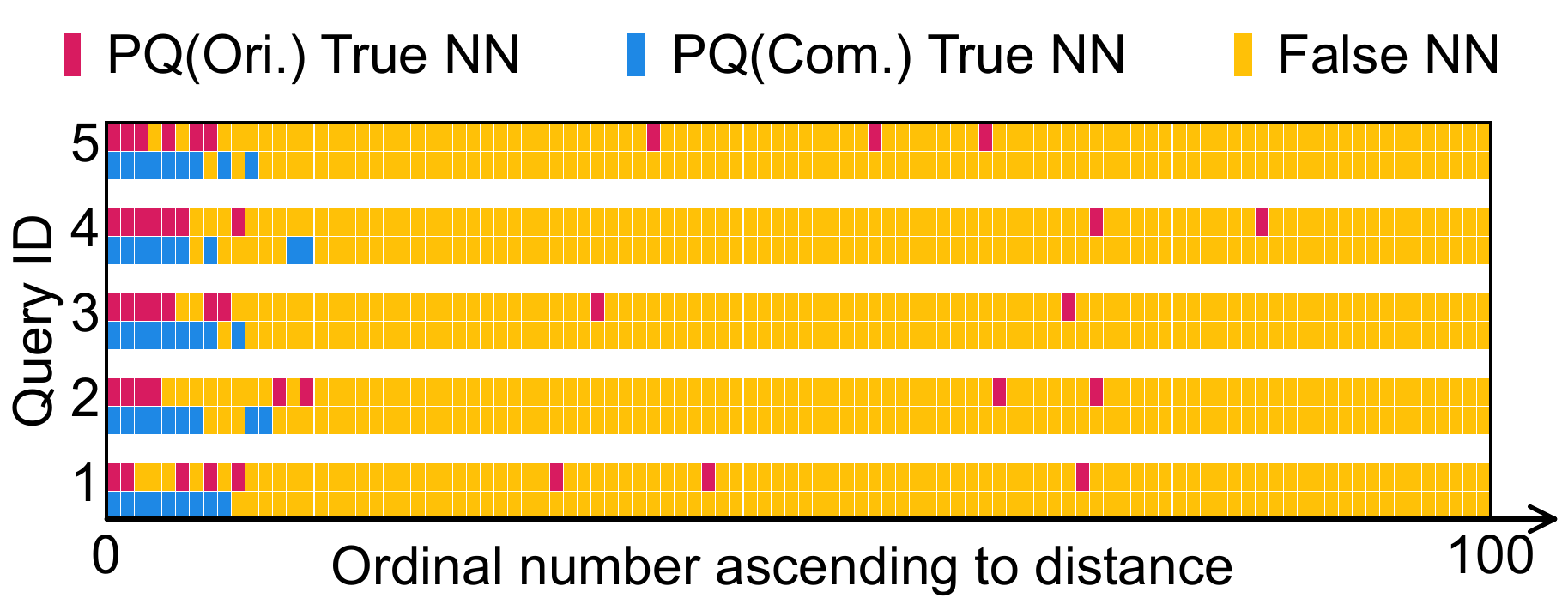}
    \vspace{-0.15in}
    \caption{Example query illustrating misranking caused by PQ error. Red and blue marks denote true nearest neighbors identified by standard PQ and hierarchical PQ, respectively, while yellow marks indicate false neighbors. Compared with standard PQ, hierarchical PQ places more true nearest neighbors near the front of the ranking.
    }
    \label{fig:heat}
    \vspace{-0.1in}
\end{figure}

By selectively improving the approximation quality of hard cases while retaining the base codebook for regular cases, HPQ reduces ranking distortion caused by extreme quantization errors. As illustrated in Figure~\ref{fig:heat}, this enables true nearest neighbors to move from the tail of the candidate queue toward earlier ranks.

\stitle{Unified Storage Design for Dual-Codebook Compression.}
For improved memory efficiency, we design a unified compression architecture that maintains two distinct codebooks (base and outlier) but stores only a single instance of compressed vectors. Specifically, for each compressed vector, we allocate one bit per subspace to designate whether the corresponding code indexes into the base or the outlier codebook. This mechanism prevents the need to store two separate sets of compressed vectors, incurring only a marginal 12.5\% overhead. As shown in Section \ref{impactofpq}, this design achieves significantly higher retrieval quality than standard quantization.

\stitle{HPQ Enhanced Two-Stage Query Strategy.}
Based on the HPQ, we propose an enhanced query strategy that optimizes both stages.

\etitle{Stage 1: HPQ Guided Greedy Search.}
The query process begins with a best-first greedy search. Unlike the naive approach, \oursys leverages the HPQ to compute approximate distances during traversal. By proactively correcting large quantization errors, HPQ generates a much higher-quality initial candidate queue.

\etitle{Stage 2: Reranking.}
We retain only the top-$\tau$ nodes from the candidate queue. Expensive disk I/O for full vector retrieval and exact distance computation is restricted exclusively to this filtered subset, significantly reducing I/O overhead without compromising recall. Note that, with HPQ, $\tau$ can be set lower than in standard PQ while achieving the same query recall. We determine $\tau$ through a warm-up calibration. A small batch of sample queries is executed to derive a baseline $\tau$ sufficient for the candidate list to meet the target recall.

\stitle{Query-Time Cost Analysis.}
Compared with standard PQ, our method adds only one extra LUT at query time for the outlier codebook. Constructing this LUT requires, for each of the $m$ subspaces, distance computations between the query sub-vector and 256 codewords, which incurs only a small preprocessing overhead compared to the distance computations and SSD accesses during search process. Its memory overhead is also minimal, requiring only $m \times 256$ float values, \ie $m \times 256 \times 4$ bytes.

\section{Implementation and Discussion}
\label{sec:implementation}
\stitle{Implementation.} We implement the designs presented in the previous sections in a dynamic ANNS system named \oursys. Furthermore, \oursys integrates three additional optimization techniques to enhance both update efficiency and query performance, which we describe in this section.


\etitle{Cache-Aware Computation.}
We optimize the existing PQ-based distance computation in FreshDiskANN and OdinANN with a cache-aware design. Our key design is to traverse nodes by subspace: distances contributed by one PQ subspace are computed for all vectors before moving to the next subspace. This flow allows the codebook entries of the current subspace to remain resident in the CPU cache, substantially reducing memory bandwidth pressure and improving cache reuse. Importantly, this optimization does not alter the computation results, but significantly accelerates distance evaluation.

\etitle{Query-Level buffer.}
To leverage our locality optimizations, we introduce a topology-only buffer that replaces the general static caching used in prior systems (\eg entry-point neighborhoods in DiskANN\cite{DiskANN} and sampled navigation nodes in Starling\cite{Starling}). 
By excluding vectors, we maximize the number of resident graph nodes. 
For better efficiency, the buffer employs a hybrid management strategy: a dynamic partition ties page lifecycle to the query context (evicting pages upon query termination to minimize cache pollution), and a small static partition pins frequently accessed nodes near the entry node to accelerate traversal initiation.

\etitle{Fine-Grained Concurrency Control.}
To support concurrent mixed workloads, we implement a lightweight, fine-grained locking mechanism in \oursys. We adopt page-level locking, restricting lock scope strictly to the target pages being operated on. Crucially, this mechanism benefits directly from our layout optimization: as related nodes are clustered into fewer pages, the number of distinct pages requiring locks during traversals is minimized. This synergy significantly reduces lock contention and waiting times, ensuring efficient concurrency even during small-batch updates, unlike Copy-on-Write schemes that suffer from heavy I/O overhead.



\stitle{Discussion.} Here we discuss the implications of our decoupled architecture across diverse hardware and algorithms, followed by an analysis of \oursys's storage overhead.

\etitle{Hardware Scalability.}
Although our current implementation targets SSDs, decoupled storage architecture provides the flexibility to place topology and raw vectors on different hardware platforms according to their distinct access patterns. For example, placing the graph topology in persistent memory exploits its byte-addressability, turning slow page-level disk I/O into fast pointer dereferences, thereby accelerating both updates and queries. Likewise, vectors can be offloaded to remote memory via RDMA, which alleviates local storage limitations by utilizing larger memory pools on remote nodes. In summary, \oursys is not a monolithic solution but a flexible architecture that adapts naturally to diverse hardware environments.

\etitle{Algorithm Scalability.}
Furthermore, the decoupled storage architecture of \oursys is potentially applicable to a broad range of graph-based indexing methods. The key reason is that many graph-based indices share the same structural property: graph topology serves as the primary data structure for traversal and connectivity maintenance during both queries and updates, while raw vectors are accessed less frequently. This suggests that the benefits of the decoupled storage architecture are not tied to a specific graph index implementation, but may extend to other graph-based structures.

\etitle{Storage Overhead.} While similarity-aware dynamic layout induces minor space amplification, it is strictly confined to the lightweight graph topology, minimizing the additional footprint (\red{1.8\%--27.8\%}). Notably, on datasets like Text2Img, \oursys actually achieves a \red{2\%} storage reduction by eliminating the internal page fragmentation caused by alignment inherent in coupled architectures, benefiting from the finer storage granularity of our decoupled design. Thus, the overall storage impact remains acceptable or even beneficial.
\section{Evaluation}
\label{EXP}

\subsection{Evaluation Setup.}

\begin{table}[t]
\centering
\caption{Benchmark datasets.} 
\vspace{-0.1in}
\begin{tabular}{ccccccc}
\hline
\textbf{Dataset} & \textbf{Dimension} & \textbf{\#Vector} & \textbf{Domain} & \textbf{Type}  \\ \hline
Deep                & 96   & 1,000,000,000        & Image              & float  \\
Sift              & 128   & 1,000,000        & Image              & float  \\ 
Text2img          & 200   & 1,000,000        & Text              & float  \\ 
Msong            & 420   & 1,000,000        & Audio      & float  \\ 
Gist                & 960   & 1,000,000        & Image              & float  \\ \hline 
\end{tabular}

\label{tab:dataset}
\vspace{-0.15in}
\end{table}

\stitle{Evaluation Platform.}
All experiments are conducted on an Aliyun ECS server (instance type: ecs.i4.8xlarge), equipped with an Intel Xeon(Ice Lake) Platinum 8369B CPU running at 2.70 GHz, 256 GB of DDR4 memory, and 2 * 3576 GB NVMe SSDs. The system runs Ubuntu 22.04, and all code is compiled with GCC 10.3.0.

\eat{
All experiments are conducted on a server equipped with an Intel Xeon Gold 5218R CPU @ 2.10 GHz, 128 GB of DDR4 memory (MICRON 3G2E1, 32 GB × 4), and 7.6 TB of SSD storage (Western Digital SN640 NVMe). The system runs CentOS Linux 7, and all code is compiled using GCC 10.3.0.}

\begin{figure*}[t]
	\centering
	\begin{minipage}[t]{\linewidth}
		\centering
		\includegraphics[width=1\linewidth]{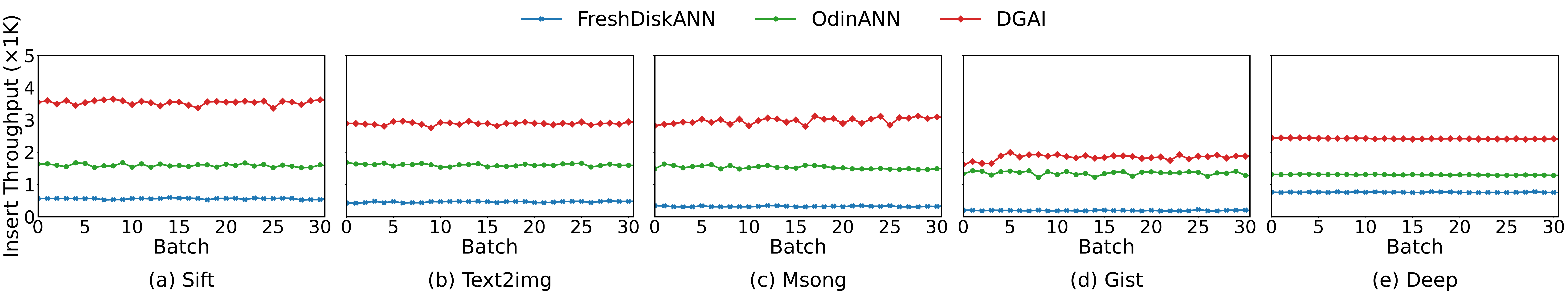}
		\label{fig:delete_perf}
	\end{minipage}
    
    \vspace{-0.3in}
	
	\begin{minipage}[t]{\linewidth}
		\centering
		\includegraphics[width=1\linewidth]{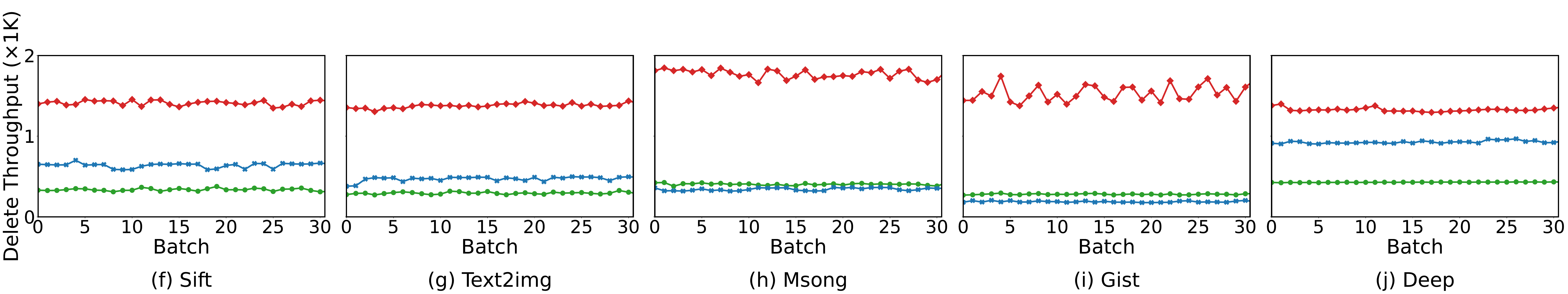}
		\label{fig:insert_perf}
	\end{minipage}
    \vspace{-0.2in}
    \caption{Comparison of update throughput for insert (top) and delete (bottom).
    }
    \label{exp:update_perf}
\end{figure*}

\stitle{Datasets.}
We evaluate our system using \red{five} public real-world datasets: \red{Deep\cite{deep1b}, Sift\cite{sift-link}, Text2img\cite{text2img-link}, Msong\cite{deep-link}, and Gist\cite{sift-link}}, covering a wide range of data dimensionalities \red{(96–960)}, scale \red{(1M-1B)} and  embedding modalities\red{(text, image and audio)}, with detailed statistics provided in Table~\ref{tab:dataset}.
Moreover, Deep1M, Deep10M, and Deep100M are subsets of the billion-scale Deep dataset, which we extract and use for scalability experiments to assess system performance under varying data sizes. Unless otherwise specified, we use Deep100M as the default scale for the Deep dataset in our evaluations.

\stitle{Compared Systems and Parameter Settings.}
Since this work focuses on storage optimization for dynamic on-disk graph-based indexes, we primarily compare \oursys against two representative dynamic graph-based ANN systems: \textbf{FreshDiskANN}~\cite{FreshDiskANN} and \textbf{OdinANN}~\cite{fast26odinann,OdinANNgithub}.
As OdinANN has already demonstrated clear performance advantages over non-graph-based dynamic systems (\eg \textbf{SPFresh}~\cite{spfresh}), we omit those baselines in our evaluation.
FreshDiskANN is a pioneering graph-based vector index designed for large-scale datasets on disk, offering low-latency queries and support for incremental updates.
OdinANN builds upon FreshDiskANN’s core architecture and introduces several key optimizations to improve insertion efficiency by implementing an in-place insertion strategy. These enhancements make OdinANN a more performant and robust solution for insertion-heavy workloads.
For a fair comparison, we apply the exact same parameter settings across all evaluated systems (our method, FreshDiskANN, and OdinANN).
Specifically, for the initial index construction, each node maintains a maximum of $R = \red{32}$ neighbors, and the insertion queue length is set to $L\_build = \red{128}$. For the deletion process, the maximum number of candidate neighbors is limited to $MAX\_C = \red{160}$. In addition, PQ is applied to compress each vector representation into \red{64} bytes.
To avoid system fluctuations, the entire procedure is repeated three times, and we report the average results.

\stitle{Initial Index.}
To accurately evaluate the effectiveness of the similarity-aware dynamic layout in \oursys, we intentionally avoid starting from an offline, globally optimized layout. Instead, we build the index from scratch and apply the same incremental insertion process used in real deployments. This ensures that the design is rigorously evaluated under realistic incremental dynamics, verifying the system's long-term stability and performance consistency.

\begin{figure}[t]
    \centering
    \includegraphics[width=\linewidth]{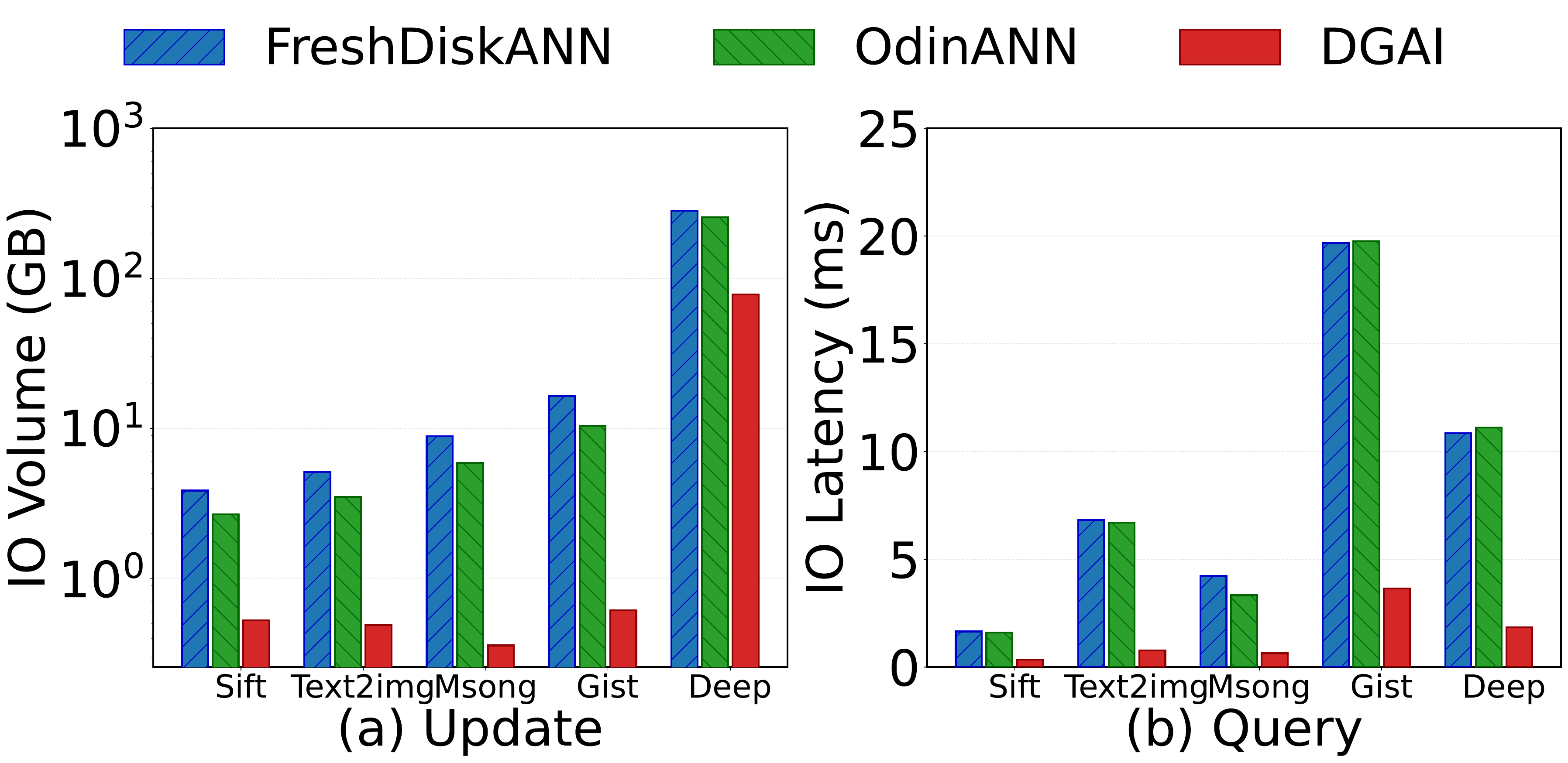}
    \vspace{-0.2in}
    \caption{Comparison of I/O in update and query.
    }
    \label{fig:update_io}
    \vspace{-0.1in}
\end{figure}

\subsection{Update Performance}\label{exp:update_}
To evaluate the update performance of each system, we first build an initial index using \red{80\%} of the dataset vectors.
We then conduct \red{32} rounds of updates for performance measurement, with each round inserting and deleting \red{1\textperthousand{}} of the total number of indexed vectors.
\oursys uses the same graph structure repair mechanism as the two baselines, 
ensuring that index quality after updates remains consistent with the baselines. Therefore, in this subsection we focus exclusively on evaluating insertion and update throughput (\ie the number of operations processed per second).

\begin{figure*}[t]
	\centering
	\begin{minipage}[t]{\linewidth}
		\centering
		\includegraphics[width=1\linewidth]{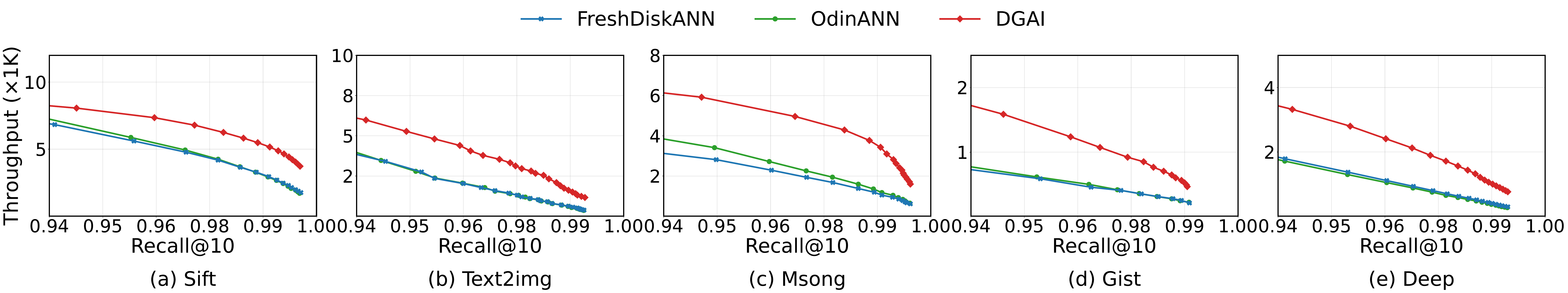}
		\label{fig:delete_perf}
	\end{minipage}
    
    \vspace{-0.3in}
	
	\begin{minipage}[t]{\linewidth}
		\centering
		\includegraphics[width=1\linewidth]{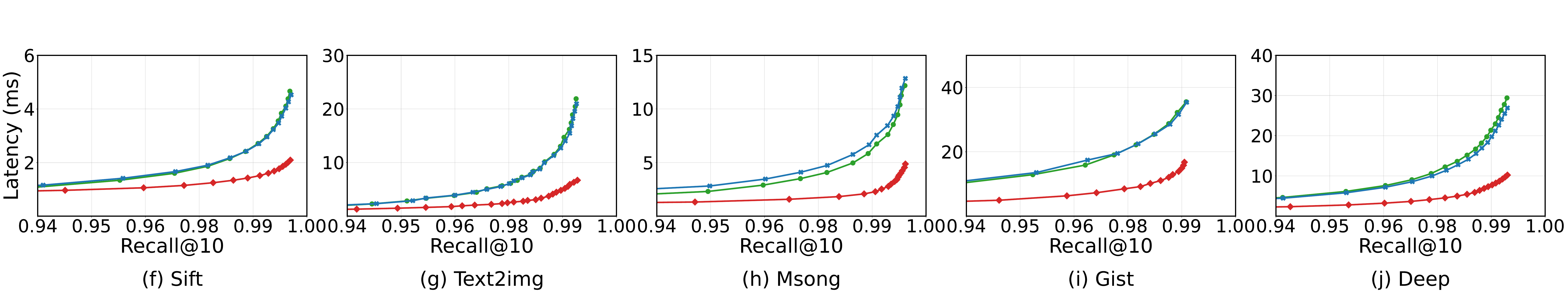}
		\label{fig:insert_perf}
	\end{minipage}
    \vspace{-0.2in}
    \caption{Comparison of query throughput (top) and latency (bottom).}
    \label{exp:query_perf}
\end{figure*}

\stitle{Insert Performance.}
Figure~\ref{exp:update_perf} (a-e) presents the insertion throughput for all evaluated systems, demonstrating that \oursys consistently delivers superior performance across all benchmarks.
Specifically, \oursys achieves a \red{3.20-8.17$\times$} speedup over FreshDiskANN and a \red{1.22-2.17$\times$} speedup over OdinANN.
We attribute this superior performance to three key designs:
(1) Our two-stage strategy enhanced by HPQ and similarity-aware dynamic layout significantly accelerates the initial phase of neighbor search for the new node (detailed in section~\ref{sec:query_perf}).
(2) The similarity-aware dynamic layout also co-locates neighbors requiring reverse edge updates within a small number of pages, reducing modified pages and I/O amplification.
(3) The in-place insertion strategy used in \oursys and OdinANN eliminates the merge overhead inherent in batch-based approaches such as FreshDiskANN, ensuring stable latency by directly updating the on-disk index.
It is also worth noting that despite maintaining a high-quality layout via insertion, \oursys incurs no visible performance penalty. The computational cost of page check is negligible compared to the substantial I/O savings it generates. Consequently, \oursys achieves state-of-the-art insertion throughput while simultaneously maintaining an optimal index layout.

\stitle{Delete Performance.}
\label{delete_performance}
Figure~\ref{exp:update_perf} (f-j) illustrates the deletion performance, where \oursys consistently and substantially outperforms both FreshDiskANN and OdinANN across all datasets.
Specifically, \oursys achieves a \red{1.43--8.16$\times$} speedup over FreshDiskANN and a \red{3.09-5.47$\times$} speedup over OdinANN.
\eat{
OdinANN performs intensive disk reorganization during the deletion phase to preserve query efficiency. This heavy overhead severely degrades deletion performance, particularly on large-scale datasets such as \red{Deep}, rendering it even slower than FreshDiskANN.}
\oursys achieves this superior efficiency through a key design: the decoupled storage architecture eliminates the need to load raw vectors during topology modifications, thereby substantially reducing I/O overhead.

To further understand the impact of our optimizations on insertion and deletion, we measure the total I/O volume over the entire update workflow.
As shown in Figure~\ref{fig:update_io} (a), \oursys reduces total I/O by \red{72.2-96.2\%} compared to FreshDiskANN and by \red{69.4-94.0\%} compared to OdinANN.
This substantial I/O reduction stems from our decoupled storage architecture and minimizing random I/O through similarity-aware dynamic layout.
These highlight the root cause of \oursys’s superior update performance: the substantial alleviation of I/O, which directly leads to higher throughput.

\subsection{Query Performance}
\label{sec:query_perf}
Figure~\ref{exp:query_perf} compares the query performance of all evaluated systems, showing that \oursys consistently achieves superior retrieval efficiency even under an update-friendly storage architecture.
Specifically, at 98\% recall, \oursys delivers \red{1.49–2.57$\times$} higher throughput than FreshDiskANN, with query latencies at only \red{38.0–65.7\%} of its levels. Compared to OdinANN, \oursys achieves \red{1.47–2.66$\times$} higher throughput while maintaining just \red{36.6–67.1\%} of its latency.

To understand the source of these gains, we analyze I/O behavior during query.
Greedy graph search typically follows a synchronous pattern: each expansion step requires loading the neighbors of the current node, computing distances, and selecting the next candidate before proceeding to the next. This sequential dependency leaves disk bandwidth underutilized. 
In contrast, our two-stage query engine decomposes the process into a fast search phase followed by candidate refinement with batched asynchronous vector I/O.
In the first stage, we leverage the similarity-aware graph layout and an ANNS co-designed buffer to access topology, reducing the I/O counts and volume to minimize the sequential part.
In the second stage, hierarchical PQ filters out redundant nodes from the ANN candidate queue.
Finally, the selected vectors are fetched in a single batched asynchronous I/O, better utilizing SSD parallelism.

Since sequential dependencies force multiple I/O requests, the I/O volume cannot faithfully reflect the actual query cost.
Therefore, we compare I/O latency during query processing, as illustrated in Figure~\ref{fig:update_io} (b).
\oursys reduces total query I/O latency by \red{78.7-88.7\%} compared to FreshDiskANN and by \red{78.1-88.5\%} against OdinANN.
These findings elucidate the root cause of the query improvements brought by \oursys.

\begin{figure}[t]
    \centering
    \includegraphics[width=0.95\linewidth]{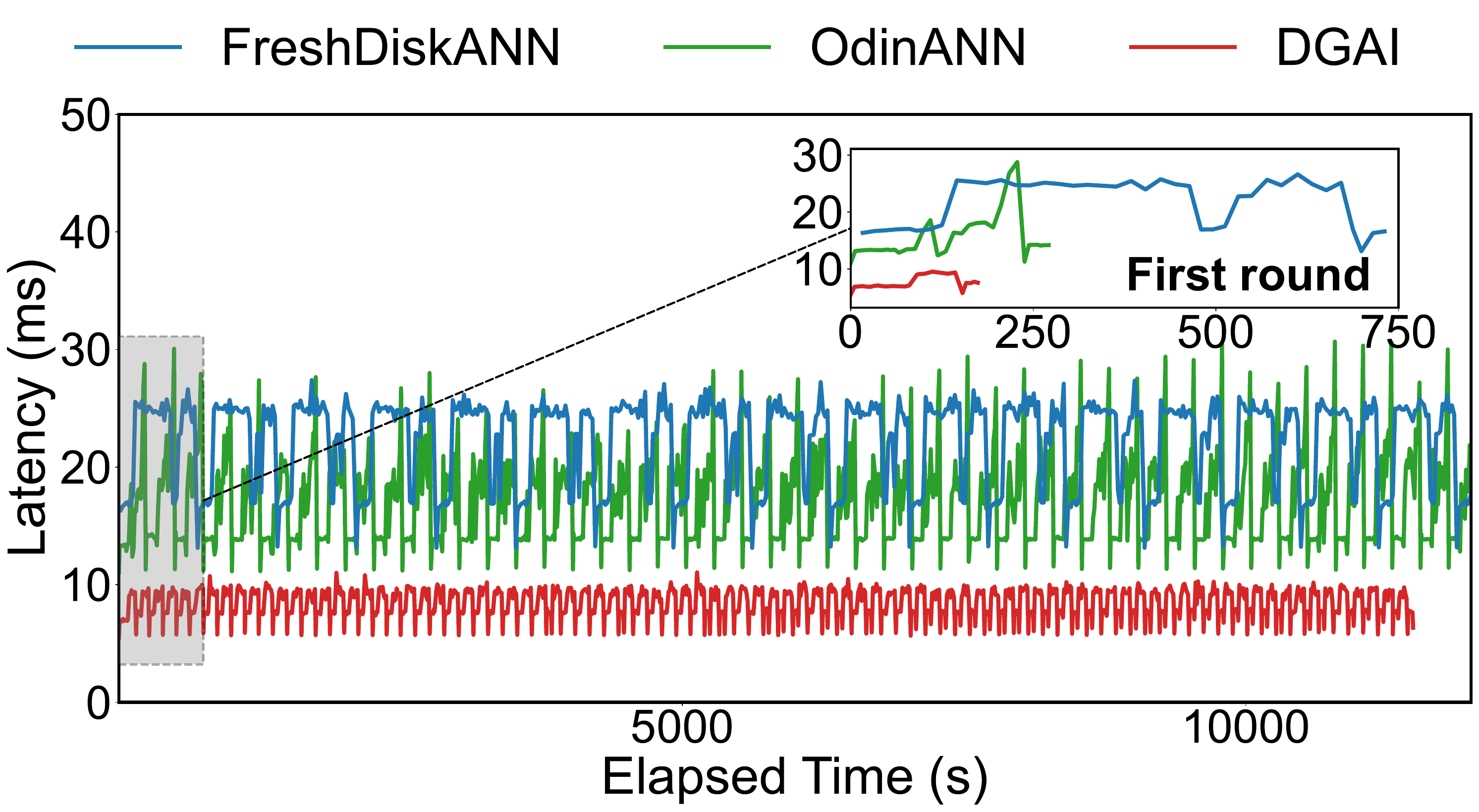}
    \vspace{-0.1in}
    \caption{P50 search latency under a concurrent mixed workload, during which the original 100M vectors in DEEP100M are continuously replaced with another 100M vectors.}
    \label{fig:mixed_workloads}
    \vspace{-0.05in}
\end{figure}

\subsection{Mixed Workload}
\eat{
To evaluate system stability under realistic dynamic conditions, we conducted a mixed-workload experiment on the \red{Deep dataset}, combining concurrent queries and updates. As shown in Figure~\ref{fig:mixed_workloads}, the most critical insight is the stark difference in latency stability.
FreshDiskANN exhibits high latency peaks during both insertion and deletion phases due to heavy resource contention. While OdinANN mitigates insertion latency via in-place updates, it still suffers from significant volatility during deletion.
In contrast, \oursys achieves exceptional stability by optimizing I/O for both updates and queries, resulting in a peak latency of only 33\% relative to OdinANN and 37\% relative to FreshDiskANN.
Furthermore, this robustness translates to superior efficiency, enabling \oursys to complete the workload \red{1.6$\times$} faster than OdinANN and \red{4.7$\times$} faster than FreshDiskANN. These results confirm that \oursys maintains distinct performance advantages and exceptional stability even in complex, high-concurrency environments.
}
To evaluate system stability under realistic dynamic conditions, we conducted a mixed-workload experiment on the \red{Deep dataset}, combining concurrent queries and updates. As shown in Figure~\ref{fig:mixed_workloads}, the most critical insight is the stark difference in latency stability among the systems. Overall, \oursys completes the entire workload \red{1.6$\times$} faster than OdinANN and \red{4.7$\times$} faster than FreshDiskANN. Furthermore, \oursys effectively suppresses latency spikes, resulting in a peak latency that is only 33\% and 37\% of that of OdinANN and FreshDiskANN, respectively.

To further analyze the root cause of this performance gap, we zoom in on the query latency variations during the first complete round of the mixed workload. As illustrated in the figure, FreshDiskANN exhibits a distinct "double-peak" pattern in its latency curve due to its batch-processing nature. The first peak is triggered by massive disk reads during the deletion phase, while the second occurs when the system intensively processes cached insertion requests. Both background operations severely interfere with foreground queries. Meanwhile, OdinANN employs a direct-insert strategy that successfully eliminates the second peak. However, its append-only write mechanism generates disk fragmentation. This exacerbates I/O amplification during deletions, leading to the sharp, unstable latency spike observed in the figure. In contrast, \oursys maintains a consistently flat and low-latency curve. By leveraging the decoupled storage architecture and the similarity-aware dynamic layout, \oursys minimizes disk access across all update stages, thereby significantly mitigating the impact of background I/O interference on foreground queries. These results confirm that \oursys delivers exceptional stability and sustained high performance even in complex, high-concurrency environments.

\begin{figure}[t]
    \centering
    \includegraphics[width=0.95\linewidth]{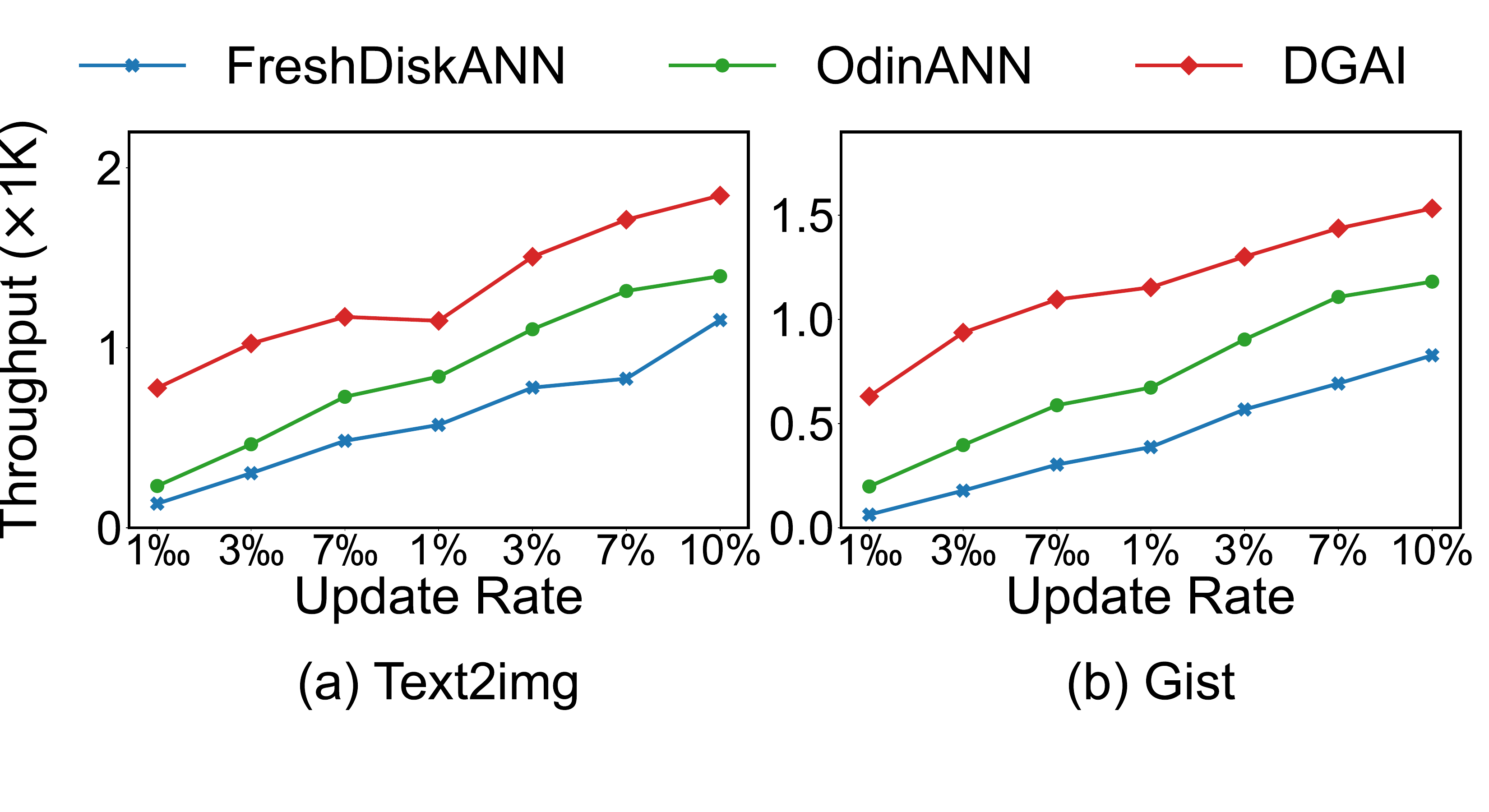}
    \vspace{-0.3in}
    \caption{Varying batch size of updates.}
    \label{fig:batch}
    
    \vspace{0.05in} 
    
    \includegraphics[width=0.95\linewidth]{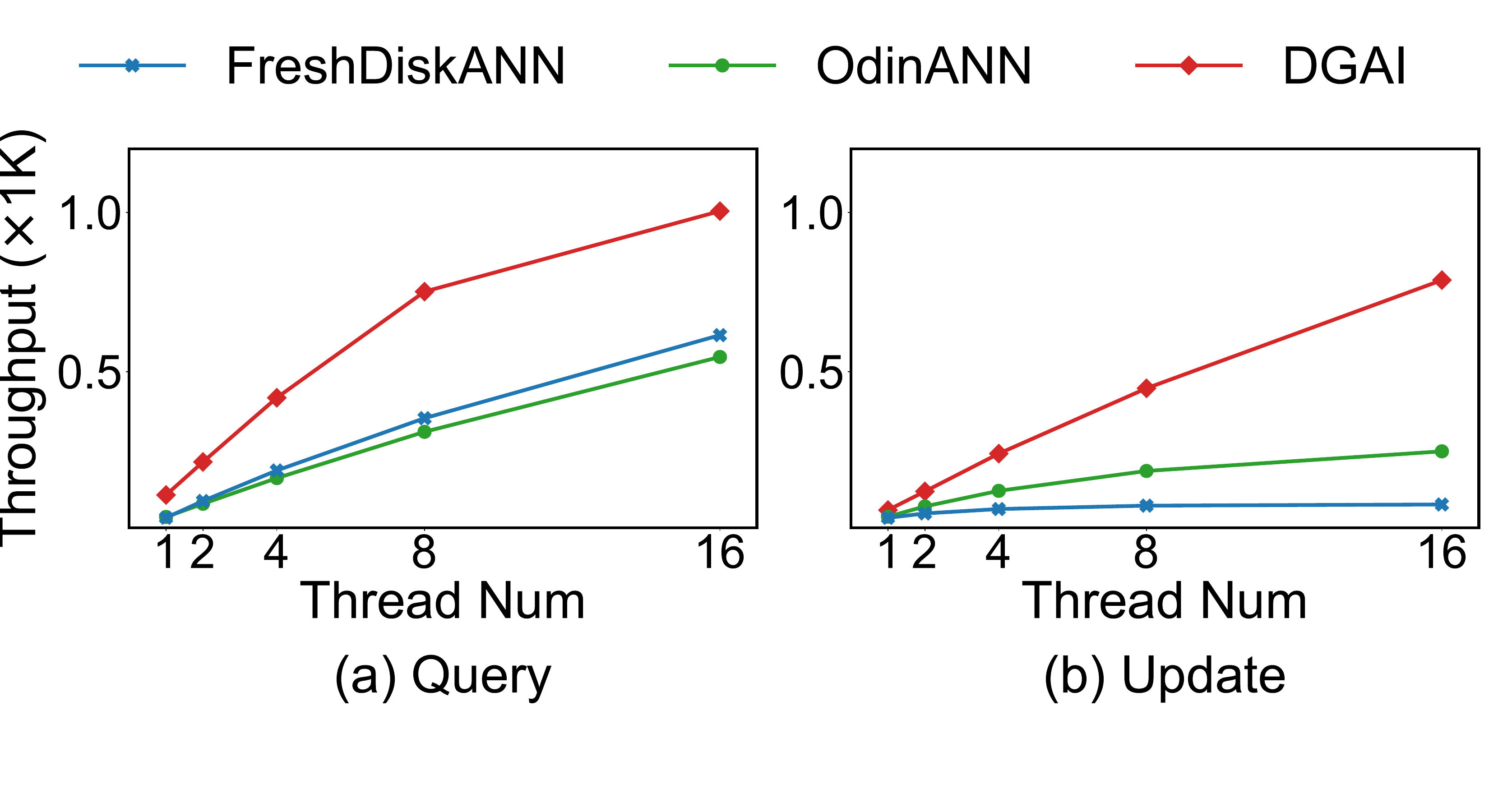}
    \vspace{-0.3in}
    \caption{Varying number of threads.}
    \label{fig:threads}
    \vspace{-0.1in}
\end{figure}



\subsection{System Scalability}
\stitle{Sensitivity to Batch Size During Updates.}
We further evaluate the impact of batch size on update performance by measuring the update throughput on the \red{Text2img and Gist} with varying batch sizes (from 1\textperthousand{} to 10\%), as shown in Figure~\ref{fig:batch}.
Since query performance is independent of batch size, update throughput serves as the primary metric.

\oursys achieves significant gains in update throughput. Specifically, it improves upon FreshDiskANN by 1.59-10.04$\times$ and OdinANN by 1.30-3.34$\times$. Despite adopting an in-place insertion mode, \oursys consistently outperforms FreshDiskANN’s batch insertion.
The advantage is especially pronounced at smaller batch sizes.
For example, on the \red{Gist} dataset, \oursys achieves a \red{1.85$\times$} speedup over FreshDiskANN with an \red{10\%} batch size, whereas this speedup widens dramatically to \red{10.04$\times$} when the batch size is reduced to 1\textperthousand{}.
Notably, even when the batch size exceeds \red{7\%}, we still maintain a clear absolute advantage.
These results confirm that our update optimizations deliver sustained advantages across batch sizes typical of real-world applications.

\stitle{Sensitivity to Thread Number.}
To evaluate system performance under parallel scenarios, we further assess its update and query scalability in a multi-threaded environment. 
We progressively increase the number of concurrent threads from \red{1 to 16} and measure the resulting throughput on \red{Gist}. 
As shown in Figure~\ref{fig:threads}, \oursys outperforms both baseline systems across all thread counts in both query and update workloads.

The scalability gap is particularly evident in update workloads. In FreshDiskANN and OdinANN, the coupled storage architecture incurs substantial redundant I/O, which causes both systems to saturate SSD bandwidth prematurely. As a result, their throughput plateaus quickly, severely limiting scalability under high parallelism. In contrast, \oursys’s decoupled storage and I/O-efficient update mechanisms avoid this bottleneck, enabling near-linear scaling with thread count.

\begin{figure}[t]
    \centering
    \includegraphics[width=0.95\linewidth]{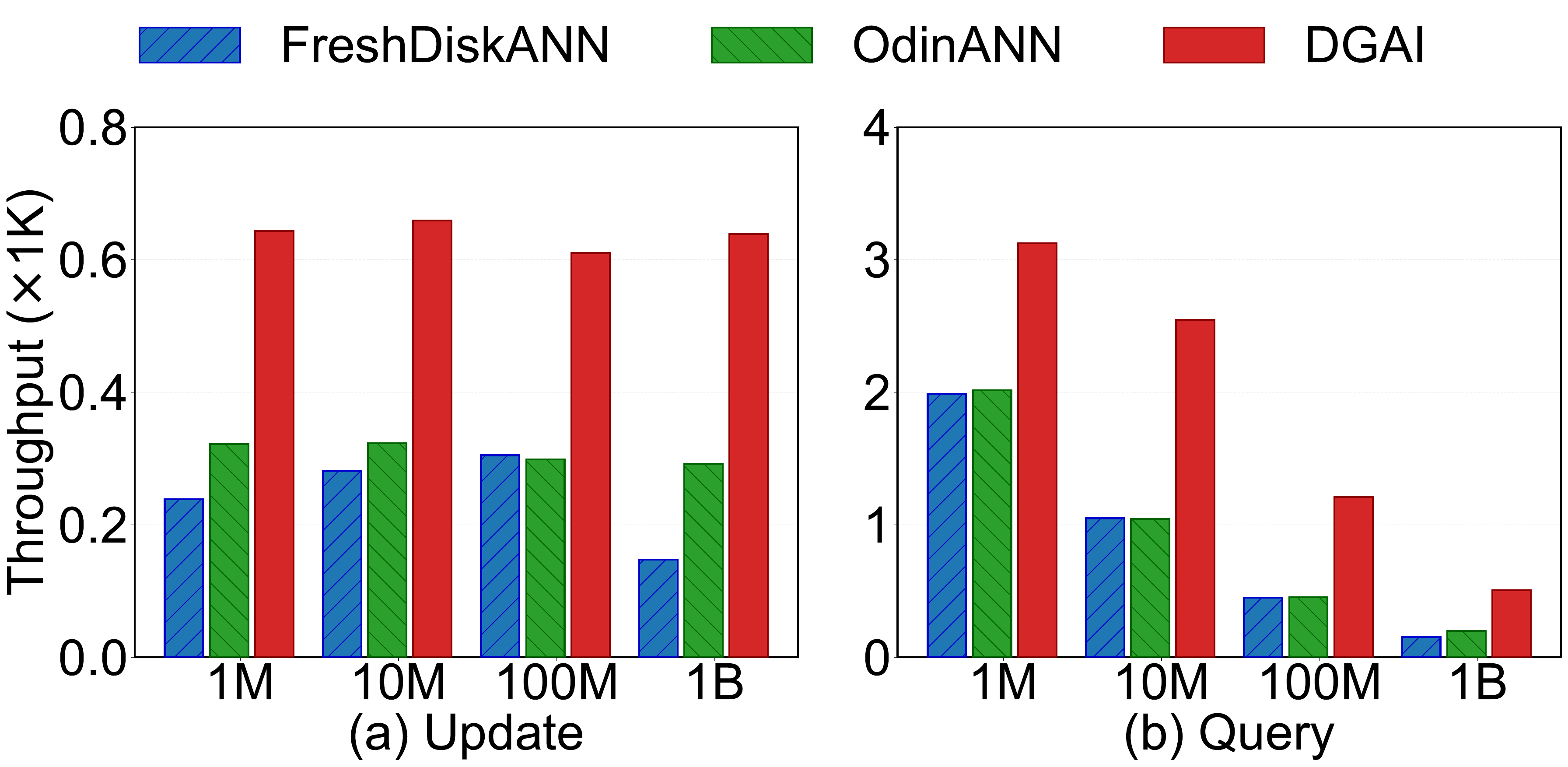}
    \vspace{-0.1in}
    \caption{Varying size of benchmark.}
    \label{fig:billionscale}
    \vspace{-0.1in}
\end{figure}

\stitle{Performance on Billion-Scale.}
We evaluate the performance of \oursys and the two baselines on the Deep dataset at multiple scales (1M, 10M, 100M, and 1B).
As shown in Figure~\ref{fig:billionscale}, \oursys consistently demonstrates superior performance across all dataset sizes.
For update throughput, \oursys consistently outperforms both baselines, achieving average improvements of \red{2.30$\times$} and \red{2.06$\times$} over FreshDiskANN and OdinANN, respectively.
For query throughput, a downward trend is observed across all systems as the dataset size increases. This is reasonable, as a larger search scope is required to maintain our target \red{98\%} recall on larger datasets, leading to increased overhead. Despite this, \oursys maintains a significant lead, outperforming the two baselines by an average of \red{2.11$\times$} and \red{1.97$\times$}.

\begin{figure}[t]
    \centering
    \includegraphics[width=\linewidth]{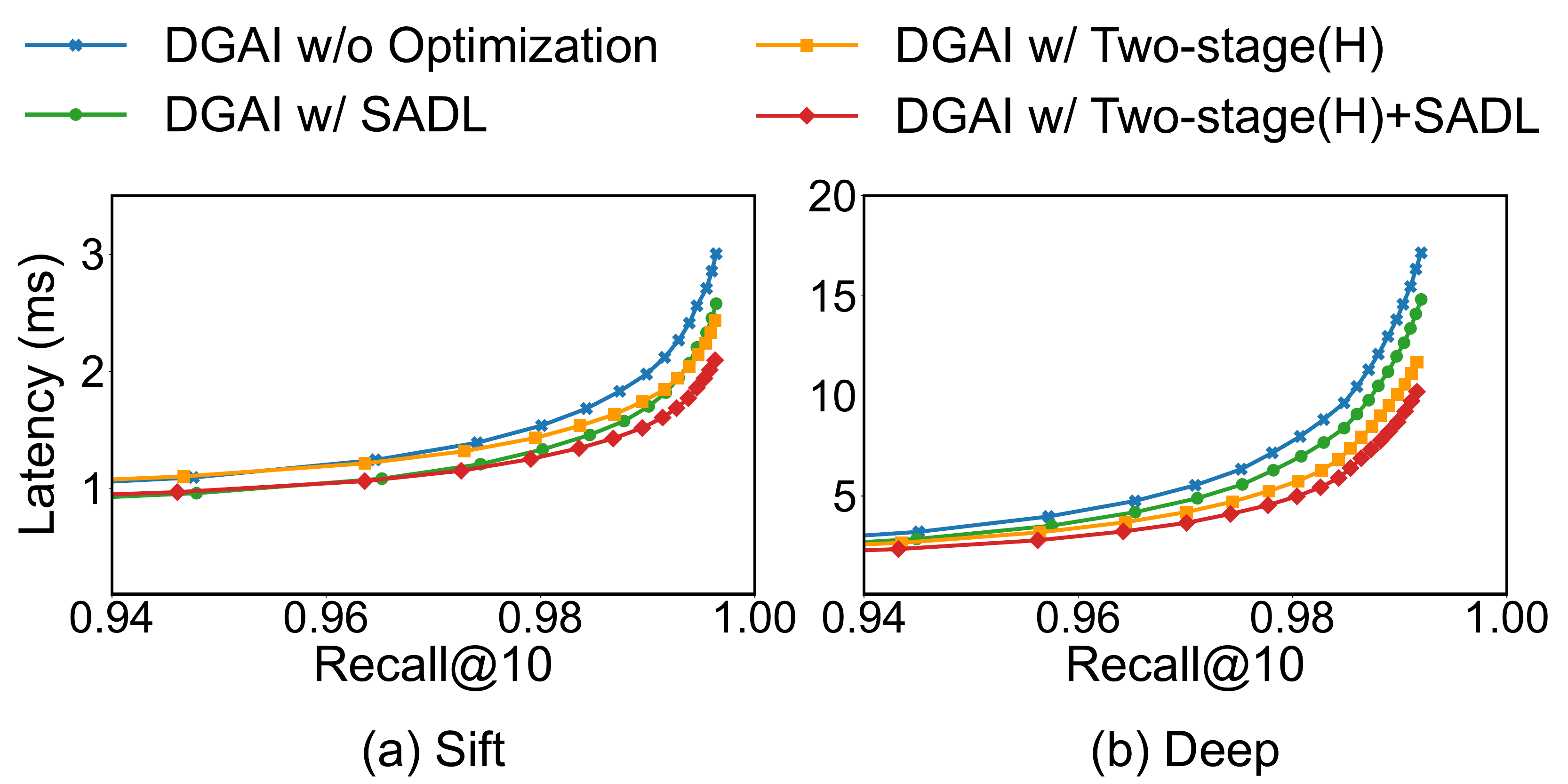}
    \vspace{-0.25in}
    \caption{Performance gain.}
    \label{fig:perform_gain}
    \vspace{-0.05in}
\end{figure}

\subsection{In-depth Analysis}
\stitle{Performance  Gain.}
To evaluate the contribution of each core design in \oursys to query performance (introduced in Section~\ref{sec:querydesign}), we incrementally incorporate the proposed components of \oursys. Each configuration is evaluated on two real-world datasets of different scales, \red{Sift and Deep}, with query latency measured at varying recall.

As shown in Figure~\ref{fig:perform_gain}, we first run \oursys without optimizations (\ie \oursys w/o Optimization) as the baseline. Incorporating the two-stage query with hierarchical PQ (\ie \oursys w/ Two-stage(H)) reduces latency by up to \red{19.0\%} on Sift and \red{28.3\%} on Deep by filtering redundant candidates before the final re-ranking.
Applying only the similarity–aware dynamic layout and co-designed buffer (\ie \oursys w/ SADL) reduces latency by up to \red{14.1\%} and \red{13.7\%} on each dataset by co-locating similar nodes within the same pages, thereby improving buffer efficiency.
When both optimizations are combined (\ie \oursys w/ Two-stage(H)+SADL), our complete query engine further achieves overall latency reductions of up to \red{30.2\%} on Sift and \red{37.5\%} on Deep. This demonstrates that the two core designs are highly complementary, addressing distinct sources of I/O overhead during query—redundant candidate processing and poor data locality—to collectively maximize retrieval efficiency.

\stitle{Impact of the Hierarchical PQ.}\label{impactofpq}
We further evaluate the effectiveness of hierarchical PQ on the Sift and Deep datasets. Specifically, we gradually increase the number of candidates ($\t$) until a recall of \red{99\%} is achieved. As shown in Table~\ref{tab:PQs}, our HPQ reaches this target with significantly fewer candidates $\tau$, resulting in a substantial total cost reduction of \red{61\%} and \red{56\%} compared to standard PQ.

To ensure a fair comparison under the same memory overhead, we further construct a standard PQ with 512 centroids, which matches the total codebook size of our hierarchical approach ($256$ for base + $256$ for outlier).
Results show that simply enlarging the codebook to 512 does not provide comparable benefits; our HPQ approach still achieves \red{27.5\%} and \red{14\%} lower cost on the two datasets.
These results indicate that HPQ strikes a more effective balance between accuracy and efficiency than brute-force codebook enlargement.

\begin{table}[t]
\centering
\caption{Impact of hierarchical PQ.}
\label{tab:PQs}
\vspace{-0.05in}
\setlength{\tabcolsep}{3pt}

\begin{tabular}{ccccc ccc}
\toprule
\multirow{2}{*}{} & \multirow{2}{*}{}
& \multicolumn{3}{c}{Sift} & \multicolumn{3}{c}{Deep} \\
\cmidrule(lr){3-5} \cmidrule(lr){6-8}
Method & Centroids & Recall & $\tau$ & I/O(KB) & Recall & $\tau$ & I/O(KB) \\
\midrule
PQ & 256 & 99.06 & 74 & 296          & 99.08 & 70 & 280 \\
HPQ & 256*2 & 99.06 & 29 & \textbf{116} & 99.08 & 31 & \textbf{124} \\
PQ & 512 & 99.06 & 40 & \uline{160}  & 99.08 & 36 & \uline{144} \\
\bottomrule
\end{tabular}

\vspace{-0.1in}
\end{table}

\section{Related Work}

\stitle{On-disk ANNS Systems.}
Approximate nearest neighbor search systems on disk are designed to handle massive vector datasets that far exceed the capacity of main memory. 
In static scenarios, where the dataset is immutable, the main objective is to maximize query performance. 
This is typically achieved by minimizing costly disk I/O operations while preserving recall, through optimizations in query execution as well as index structures and data layouts~\cite{DiskANN,Starling}.
In contrast, dynamic systems face the more complex challenge of efficiently handling frequent updates.
This requires a careful balance between sustaining high query performance and minimizing update latency.
To address this trade-off, several innovative solutions have emerged.
For example, FreshDiskANN~\cite{FreshDiskANN} is a pioneering graph-based index that introduced support for dynamic operations. 
Instead of incurring the prohibitive cost of a full index rebuild, it incrementally adjusts the graph in response to updates. This incremental maintenance strategy preserves high index quality and superior query performance, establishing a foundational approach for subsequent dynamic graph-based ANN systems~\cite{fast26odinann,liuwolverine,pgvector,IP-DiskANN}.
Greator\cite{Greator} adopts a separately maintained topology replica for fast in-neighbors lookup. However, this is essentially a replicated design rather than a truly decoupled one, since its entire update process is still carried out on the original architecture where topology and vectors remain coupled. As a result, Greator still suffers from I/O overhead during updates.
Parallel to graph-based approaches, cluster-based methods like SPFresh~\cite{spfresh} employ the LIRE strategy to dynamically maintain cluster quality. However, a fundamental limitation of such cluster-based algorithms is their reliance on fetching massive amounts of candidate points from disk for exact distance computation (i.e., reranking) during the search phase. This process generates overwhelming disk I/O requests that quickly saturate I/O bandwidth. 
    
\stitle{Vector Database.}
To meet the surging demand for managing high-dimensional vector data, a proliferation of vector database systems has emerged~\cite{AnalyticDB-V, PASE, SingleStore-V, VBASE, Milvus, Pinecone, vexless, Vearch, zilliz}. 
Initially, traditional relational databases, such as PostgreSQL~\cite{pgvector} and MySQL~\cite{mysql}, integrated vector operations as extensions, allowing users to manage vectors within a familiar relational framework. However, since relational storage engines and execution models were not natively designed for dense vector computations, they often fail to fully exploit the mathematical properties of vectors, leading to suboptimal performance.
Consequently, purpose-built vector databases, such as Milvus~\cite{Milvus}, Qdrant, and Weaviate, have dominated the market. By co-designing storage layouts with state-of-the-art ANNS algorithms (e.g., Faiss~\cite{Faiss}) and leveraging hardware-accelerated execution, these specialized systems deliver superior throughput and scalability, particularly in large-scale scenarios.

\section{Conclusion}

We propose a dynamic on-disk ANNS system named \oursys, which employs a decoupled storage architecture that separates vector data from the graph index to minimize I/O overhead during updates. Furthermore, we design a similarity-aware dynamic layout and a two-stage query mechanism enhanced by hierarchical PQ, substantially improving query performance under the decoupled storage architecture.
Experiments show that \oursys significantly outperforms state-of-the-art systems, achieving higher update throughput and lower query latency under dynamic workloads.


\bibliographystyle{ACM-Reference-Format}
\bibliography{sample}

\end{document}